\input harvmac
\input epsf

\newcount\figno
\figno=0
\def\fig#1#2#3{
\par\begingroup\parindent=0pt\leftskip=1cm\rightskip=1cm\parindent=0pt
\baselineskip=11pt \global\advance\figno by 1 \midinsert
\epsfxsize=#3 \centerline{\epsfbox{#2}} \vskip 12pt {\bf Figure
\the\figno:} #1\par
\endinsert\endgroup\par
}
\def\figlabel#1{\xdef#1{\the\figno}}

\def\journal#1&#2(#3){\unskip, \sl #1\ \bf #2 \rm(19#3) }
\def\andjournal#1&#2(#3){\sl #1~\bf #2 \rm (19#3) }

\def\ie{{\it i.e.}}
\def\eg{{\it e.g.}}

\noblackbox
%


\def\unlockat{\catcode`\@=11}
\def\lockat{\catcode`\@=12}

\unlockat

\def\newsec#1{\global\advance\secno by1\message{(\the\secno. #1)}
\global\subsecno=0\global\subsubsecno=0\eqnres@t\noindent
{\bf\the\secno. #1}
\writetoca{{\secsym} {#1}}\par\nobreak\medskip\nobreak}
\global\newcount\subsecno \global\subsecno=0
\def\subsec#1{\global\advance\subsecno
by1\message{(\secsym\the\subsecno. #1)}
\ifnum\lastpenalty>9000\else\bigbreak\fi\global\subsubsecno=0
\noindent{\it\secsym\the\subsecno. #1}
\writetoca{\string\quad {\secsym\the\subsecno.} {#1}}
\par\nobreak\medskip\nobreak}
\global\newcount\subsubsecno \global\subsubsecno=0
\def\subsubsec#1{\global\advance\subsubsecno by1
\message{(\secsym\the\subsecno.\the\subsubsecno. #1)}
\ifnum\lastpenalty>9000\else\bigbreak\fi
\noindent\quad{\secsym\the\subsecno.\the\subsubsecno.}{#1}
\writetoca{\string\qquad{\secsym\the\subsecno.\the\subsubsecno.}{#1}}
\par\nobreak\medskip\nobreak}

\def\subsubseclab#1{\DefWarn#1\xdef
#1{\noexpand\hyperref{}{subsubsection}%
{\secsym\the\subsecno.\the\subsubsecno}%
{\secsym\the\subsecno.\the\subsubsecno}}%
\writedef{#1\leftbracket#1}\wrlabeL{#1=#1}}
\lockat

\def\ie{{\it i.e.}}
\def\eg{{\it e.g.}}

\def\CM {{\cal M}}
\def\CN {{\cal N}}

\def\CZ {{\cal Z }}
\def\CS {{\cal S }}
\def\ch{{\rm ch}}

\font\manual=manfnt \def\dbend{\lower3.5pt\hbox{\manual\char127}}

\def\IZ{\relax\ifmmode\mathchoice
{\hbox{\cmss Z\kern-.4em Z}}{\hbox{\cmss Z\kern-.4em Z}}
{\lower.9pt\hbox{\cmsss Z\kern-.4em Z}}
{\lower1.2pt\hbox{\cmsss Z\kern-.4em Z}}\else{\cmss Z\kern-.4em
Z}\fi}
\def\half{{1\over 2}}

\def\CM {{\cal M}}
\def\CN {{\cal N}}

\def\CZ {{\cal Z }}
\def\CS {{\cal S }}
\def\ch{{\rm ch}}


\def\IZ{\relax\ifmmode\mathchoice
{\hbox{\cmss Z\kern-.4em Z}}{\hbox{\cmss Z\kern-.4em Z}}
{\lower.9pt\hbox{\cmsss Z\kern-.4em Z}}
{\lower1.2pt\hbox{\cmsss Z\kern-.4em Z}}\else{\cmss Z\kern-.4em
Z}\fi}
\def\IB{\relax{\rm I\kern-.18em B}}
\def\IC{{\relax\hbox{$\inbar\kern-.3em{\rm C}$}}}
\def\ID{\relax{\rm I\kern-.18em D}}
\def\IE{\relax{\rm I\kern-.18em E}}
\def\IF{\relax{\rm I\kern-.18em F}}
\def\IG{\relax\hbox{$\inbar\kern-.3em{\rm G}$}}
\def\IGa{\relax\hbox{${\rm I}\kern-.18em\Gamma$}}
\def\IH{\relax{\rm I\kern-.18em H}}
\def\II{\relax{\rm I\kern-.18em I}}
\def\IK{\relax{\rm I\kern-.18em K}}
\def\IP{\relax{\rm I\kern-.18em P}}
\def\IQ{\relax\hbox{$\inbar\kern-.3em{\rm Q}$}}

\def\inbar{\,\vrule height1.5ex width.4pt depth0pt}

\font\cmss=cmss10 \font\cmsss=cmss10 at 7pt
\def\IR{\relax{\rm I\kern-.18em R}}

%
%

\def\makeblankbox#1#2{\hbox{\lower\dp0\vbox{\hidehrule{#1}{#2}%
   \kern -#1
   \hbox to \wd0{\hidevrule{#1}{#2}%
      \raise\ht0\vbox to #1{}
      \lower\dp0\vtop to #1{}
      \hfil\hidevrule{#2}{#1}}%
   \kern-#1\hidehrule{#2}{#1}}}%
}%
\def\hidehrule#1#2{\kern-#1\hrule height#1 depth#2 \kern-#2}%
\def\hidevrule#1#2{\kern-#1{\dimen0=#1\advance\dimen0 by #2\vrule
    width\dimen0}\kern-#2}%
\def\openbox{\ht0=1.2mm \dp0=1.2mm \wd0=2.4mm  \raise 2.75pt
\makeblankbox {.25pt} {.25pt}  }

\def\bun#1/#2{\leavevmode
   \kern.1em \raise .5ex \hbox{\the\scriptfont0 #1}%
   \kern-.1em $/$%
   \kern-.15em \lower .25ex \hbox{\the\scriptfont0 #2}%
}

\def\opensquare{\ht0=3.4mm \dp0=3.4mm \wd0=6.8mm  \raise 2.7pt
\makeblankbox {.25pt} {.25pt}  }


\def\sector#1#2{\ {\scriptstyle #1}\hskip 1mm
\mathop{\opensquare}\limits_{\lower 1mm\hbox{$\scriptstyle#2$}}\hskip 1mm}

\def\tsector#1#2{\ {\scriptstyle #1}\hskip 1mm
\mathop{\opensquare}\limits_{\lower 1mm\hbox{$\scriptstyle#2$}}^\sim\hskip 1mm}


\def\inbar{\,\vrule height1.5ex width.4pt depth0pt}

\font\cmss=cmss10 \font\cmsss=cmss10 at 7pt
\def\IR{\relax{\rm I\kern-.18em R}}


\def\sst{\scriptscriptstyle}

\def\frac#1#2{{#1\over#2}}

\def\half{\frac12}

\def\d{\partial}

\def\inbar{\,\vrule height1.5ex width.4pt depth0pt}
\def\IC{\relax\hbox{$\inbar\kern-.3em{\rm C}$}}
\def\IR{\relax{\rm I\kern-.18em R}}
\def\IP{\relax{\rm I\kern-.18em P}}

%
%
\catcode`\@=11
\def\slash#1{\mathord{\mathpalette\c@ncel{#1}}}
\overfullrule=0pt

\def\II{{\cal I}}

\def\RR{{\cal R}}

\def\underrel#1\over#2{\mathrel{\mathop{\kern\z@#1}\limits_{#2}}}

\catcode`\@=12


%

\def\cosh{{\rm cosh}}

\def\exp{{\rm exp}}
\def\sh{{\rm sh}}
\def\ch{{\rm ch}}


\def\ceff{c_{\rm eff}}

\def\sst{\scriptscriptstyle}

\def\frac#1#2{{#1\over#2}}

\def\half{\frac12}

\def\d{\partial}

\def\inbar{\,\vrule height1.5ex width.4pt depth0pt}
\def\IC{\relax\hbox{$\inbar\kern-.3em{\rm C}$}}
\def\IR{\relax{\rm I\kern-.18em R}}
\def\IP{\relax{\rm I\kern-.18em P}}

%
%

%
\catcode`\@=11
\def\slash#1{\mathord{\mathpalette\c@ncel{#1}}}
\overfullrule=0pt

\def\II{{\cal I}}

\def\RR{{\cal R}}

\def\underrel#1\over#2{\mathrel{\mathop{\kern\z@#1}\limits_{#2}}}

\catcode`\@=12


%

\def \cosh{{\rm cosh}}

\def\exp{{\rm exp}}
\def\sh{{\rm sinh}}
\def\ch{{\rm cosh}}


\def\({\left(}
\def\){\right)}
\def\[{\left[}
\def\]{\right]}

\lref\AharonyTI{
O.~Aharony, S.~S.~Gubser, J.~M.~Maldacena, H.~Ooguri and Y.~Oz,
``Large N field theories, string theory and gravity,''
Phys.\ Rept.\  {\bf 323}, 183 (2000)
[arXiv:hep-th/9905111].
}

\lref\BrownNW{
J.~D.~Brown and M.~Henneaux,
``Central Charges In The Canonical Realization Of Asymptotic Symmetries: An
Example From Three-Dimensional Gravity,''
Commun.\ Math.\ Phys.\  {\bf 104}, 207 (1986).
}

\lref\CallanAT{
C.~G.~.~Callan, J.~A.~Harvey and A.~Strominger,
``Supersymmetric string solitons,''
arXiv:hep-th/9112030.
}

\lref\KutasovXU{
D.~Kutasov and N.~Seiberg,
``More comments on string theory on AdS(3),''
JHEP {\bf 9904}, 008 (1999)
[arXiv:hep-th/9903219].
}

\lref\StromingerEQ{
A.~Strominger,
``Black hole entropy from near-horizon microstates,''
JHEP {\bf 9802}, 009 (1998)
[arXiv:hep-th/9712251].
}

\lref\GiveonNS{
A.~Giveon, D.~Kutasov and N.~Seiberg,
``Comments on string theory on AdS(3),''
Adv.\ Theor.\ Math.\ Phys.\  {\bf 2}, 733 (1998)
[arXiv:hep-th/9806194].
}

\lref\SeibergXZ{
N.~Seiberg and E.~Witten,
``The D1/D5 system and singular CFT,''
JHEP {\bf 9904}, 017 (1999)
[arXiv:hep-th/9903224].
}

\lref\KutasovUA{
D.~Kutasov and N.~Seiberg,
``Noncritical Superstrings,''
Phys.\ Lett.\ B {\bf 251}, 67 (1990).
}

\lref\GiveonZM{
A.~Giveon, D.~Kutasov and O.~Pelc,
``Holography for non-critical superstrings,''
JHEP {\bf 9910}, 035 (1999)
[arXiv:hep-th/9907178].
}

\lref\SusskindWS{
L.~Susskind,
``Some speculations about black hole entropy in string theory,''
arXiv:hep-th/9309145.
}

\lref\HorowitzNW{
G.~T.~Horowitz and J.~Polchinski,
``A correspondence principle for black holes and strings,''
Phys.\ Rev.\ D {\bf 55}, 6189 (1997)
[arXiv:hep-th/9612146].
}

\lref\MaldacenaHW{
J.~M.~Maldacena and H.~Ooguri,
``Strings in AdS(3) and SL(2,R) WZW model. I,''
J.\ Math.\ Phys.\  {\bf 42}, 2929 (2001)
[arXiv:hep-th/0001053].
}

\lref\KutasovSV{
D.~Kutasov and N.~Seiberg,
``Number Of Degrees Of Freedom, Density Of States And Tachyons In String
Theory And Cft,''
Nucl.\ Phys.\ B {\bf 358}, 600 (1991).
}

\lref\KutasovZH{
D.~Kutasov, F.~Larsen and R.~G.~Leigh,
``String theory in magnetic monopole backgrounds,''
Nucl.\ Phys.\ B {\bf 550}, 183 (1999)
[arXiv:hep-th/9812027].
}

\lref\GiveonUP{
A.~Giveon and D.~Kutasov,
``Notes on AdS(3),''
Nucl.\ Phys.\ B {\bf 621}, 303 (2002)
[arXiv:hep-th/0106004].
}

\lref\KutasovPV{
D.~Kutasov,
``Some properties of (non)critical strings,''
arXiv:hep-th/9110041.
}

\lref\GiveonPX{
A.~Giveon and D.~Kutasov,
``Little string theory in a double scaling limit,''
JHEP {\bf 9910}, 034 (1999)
[arXiv:hep-th/9909110].
}

\lref\TeschnerRV{
J.~Teschner,
``Liouville theory revisited,''
Class.\ Quant.\ Grav.\  {\bf 18}, R153 (2001)
[arXiv:hep-th/0104158].
}

\lref\NakayamaVK{
Y.~Nakayama,
``Liouville field theory: A decade after the revolution,''
Int.\ J.\ Mod.\ Phys.\ A {\bf 19}, 2771 (2004)
[arXiv:hep-th/0402009].
}

\lref\AharonyUB{
O.~Aharony, M.~Berkooz, D.~Kutasov and N.~Seiberg,
``Linear dilatons, NS5-branes and holography,''
JHEP {\bf 9810}, 004 (1998)
[arXiv:hep-th/9808149].
}

\lref\KutasovJP{
D.~Kutasov and D.~A.~Sahakyan,
``Comments on the thermodynamics of little string theory,''
JHEP {\bf 0102}, 021 (2001)
[arXiv:hep-th/0012258].
}

\lref\AharonyXN{
O.~Aharony, A.~Giveon and D.~Kutasov,
``LSZ in LST,''
Nucl.\ Phys.\ B {\bf 691}, 3 (2004)
[arXiv:hep-th/0404016].
}

\lref\KazakovPM{
V.~Kazakov, I.~K.~Kostov and D.~Kutasov,
``A matrix model for the two-dimensional black hole,''
Nucl.\ Phys.\ B {\bf 622}, 141 (2002)
[arXiv:hep-th/0101011].
}

\lref\KarczmarekBW{
J.~L.~Karczmarek, J.~Maldacena and A.~Strominger,
``Black hole non-formation in the matrix model,''
arXiv:hep-th/0411174.
}

\lref\WittenYR{
  E.~Witten,
  ``On string theory and black holes,''
  Phys.\ Rev.\ D {\bf 44}, 314 (1991).
}

\lref\GiveonFU{
  A.~Giveon, M.~Porrati and E.~Rabinovici,
  ``Target space duality in string theory,''
  Phys.\ Rept.\  {\bf 244}, 77 (1994)
  [arXiv:hep-th/9401139].
}

\lref\MandalTZ{
  G.~Mandal, A.~M.~Sengupta and S.~R.~Wadia,
  ``Classical solutions of two-dimensional string theory,''
  Mod.\ Phys.\ Lett.\ A {\bf 6}, 1685 (1991).
}

\lref\ElitzurCB{
  S.~Elitzur, A.~Forge and E.~Rabinovici,
  ``Some global aspects of string compactifications,''
  Nucl.\ Phys.\ B {\bf 359}, 581 (1991).
}

\lref\GiveonSY{
  A.~Giveon,
  ``Target space duality and stringy black holes,''
  Mod.\ Phys.\ Lett.\ A {\bf 6}, 2843 (1991).
}

\lref\DijkgraafBA{
  R.~Dijkgraaf, H.~Verlinde and E.~Verlinde,
  ``String propagation in a black hole geometry,''
  Nucl.\ Phys.\ B {\bf 371}, 269 (1992).
}

\lref\McGuiganQP{
  M.~D.~McGuigan, C.~R.~Nappi and S.~A.~Yost,
  ``Charged black holes in two-dimensional string theory,''
  Nucl.\ Phys.\ B {\bf 375}, 421 (1992)
  [arXiv:hep-th/9111038].
}

\lref\NappiAS{
  C.~R.~Nappi and A.~Pasquinucci,
  ``Thermodynamics of two-dimensional black holes,''
  Mod.\ Phys.\ Lett.\ A {\bf 7}, 3337 (1992)
  [arXiv:gr-qc/9208002].
}

\lref\GiveonRW{
  A.~Giveon, A.~Konechny, E.~Rabinovici and A.~Sever,
  ``On thermodynamical properties of some coset CFT backgrounds,''
  JHEP {\bf 0407}, 076 (2004)
  [arXiv:hep-th/0406131].
}

\lref\GiveonGE{
  A.~Giveon, E.~Rabinovici and A.~Sever,
  ``Beyond the singularity of the 2-D charged black hole,''
  JHEP {\bf 0307}, 055 (2003)
  [arXiv:hep-th/0305140].
}

\lref\JohnsonJW{
  C.~V.~Johnson,
  ``Exact models of extremal dyonic 4-D black hole solutions of heterotic
  string theory,''
  Phys.\ Rev.\ D {\bf 50}, 4032 (1994)
  [arXiv:hep-th/9403192].
}

\lref\GiveonJG{
  A.~Giveon and M.~Rocek,
  ``Supersymmetric string vacua on AdS(3) x N,''
  JHEP {\bf 9904}, 019 (1999)
  [arXiv:hep-th/9904024].
}

\lref\BerensteinGJ{
  D.~Berenstein and R.~G.~Leigh,
  ``Spacetime supersymmetry in AdS(3) backgrounds,''
  Phys.\ Lett.\ B {\bf 458}, 297 (1999)
  [arXiv:hep-th/9904040].
}

\lref\GiveonZZ{
  A.~Giveon and A.~Sever,
  ``Strings in a 2-d extremal black hole,''
  arXiv:hep-th/0412294.
}

\lref\ArgurioTB{
  R.~Argurio, A.~Giveon and A.~Shomer,
  ``Superstrings on AdS(3) and symmetric products,''
  JHEP {\bf 0012}, 003 (2000)
  [arXiv:hep-th/0009242].
}

\lref\HorneGN{
  J.~H.~Horne and G.~T.~Horowitz,
  ``Exact black string solutions in three-dimensions,''
  Nucl.\ Phys.\ B {\bf 368}, 444 (1992)
  [arXiv:hep-th/9108001].
}

\lref\bars{I.~Bars and K.~Sfetsos, ``Conformally exact metric and
dilaton in string theory on curved space-time,'' Phys.\ Rev.\ D
{\bf 46}, 4510 (1992) [arXiv:hep-th/9206006]; A.~A.~Tseytlin,
``Conformal sigma models corresponding to gauged
Wess-Zumino-Witten theories,'' Nucl.\ Phys.\ B {\bf 411}, 509
(1994) [arXiv:hep-th/9302083].}

\lref\HawkingFD{
  S.~W.~Hawking and G.~T.~Horowitz,
  ``The Gravitational Hamiltonian, action, entropy and surface terms,''
  Class.\ Quant.\ Grav.\  {\bf 13}, 1487 (1996)
  [arXiv:gr-qc/9501014].
}

\lref\ElitzurVW{S.~Elitzur, A.~Giveon and E.~Rabinovici,
``Removing singularities,'' JHEP {\bf 0301}, 017 (2003)
[arXiv:hep-th/0212242].}

\lref\AharonyTH{
O.~Aharony, M.~Berkooz, S.~Kachru, N.~Seiberg and E.~Silverstein,
``Matrix description of interacting theories in six dimensions,''
Adv.\ Theor.\ Math.\ Phys.\  {\bf 1}, 148 (1998)
[arXiv:hep-th/9707079].
}

\lref\GanorJX{
O.~J.~Ganor and S.~Sethi,
``New perspectives on Yang-Mills theories with sixteen supersymmetries,''
JHEP {\bf 9801}, 007 (1998)
[arXiv:hep-th/9712071].
}

\lref\KlemmDF{
A.~Klemm and M.~G.~Schmidt,
``Orbifolds By Cyclic Permutations Of Tensor Product Conformal Field
Theories,''
Phys.\ Lett.\ B {\bf 245}, 53 (1990).
}

\lref\FuchsVU{
J.~Fuchs, A.~Klemm and M.~G.~Schmidt,
``Orbifolds by cyclic permutations in Gepner type superstrings and in the
corresponding Calabi-Yau manifolds,''
Annals Phys.\  {\bf 214}, 221 (1992).
}

\lref\AchucarroVZ{
A.~Achucarro and P.~K.~Townsend,
``A Chern-Simons Action For Three-Dimensional Anti-De Sitter Supergravity
Theories,''
Phys.\ Lett.\ B {\bf 180}, 89 (1986).
}

\lref\WittenHC{
E.~Witten,
``(2+1)-Dimensional Gravity As An Exactly Soluble System,''
Nucl.\ Phys.\ B {\bf 311}, 46 (1988).
}

\lref\CoussaertZP{
O.~Coussaert, M.~Henneaux and P.~van Driel,
``The Asymptotic dynamics of three-dimensional Einstein gravity with a
negative cosmological constant,''
Class.\ Quant.\ Grav.\  {\bf 12}, 2961 (1995)
[arXiv:gr-qc/9506019].
}

\lref\ChenSI{
Y.~j.~Chen,
``Quantum Liouville theory and BTZ black hole entropy,''
Class.\ Quant.\ Grav.\  {\bf 21}, 1153 (2004)
[arXiv:hep-th/0310234].
}

\lref\FidkowskiFC{
L.~Fidkowski and S.~Shenker,
``D-brane instability as a large N phase transition,''
arXiv:hep-th/0406086.
}

\lref\BirminghamJT{
D.~Birmingham, I.~Sachs and S.~Sen,
``Entropy of three-dimensional black holes in string theory,''
Phys.\ Lett.\ B {\bf 424}, 275 (1998)
[arXiv:hep-th/9801019].
}

\lref\CarlipZN{
S.~Carlip,
``Conformal field theory, (2+1)-dimensional gravity, and the BTZ black
hole,''
arXiv:gr-qc/0503022.
}

\lref\BirminghamDT{
D.~Birmingham, I.~Sachs and S.~Sen,
``Exact results for the BTZ black hole,''
Int.\ J.\ Mod.\ Phys.\ D {\bf 10}, 833 (2001)
[arXiv:hep-th/0102155].
}

\Title{
}
{\vbox{
\centerline{Phases of Quantum Gravity in}
\bigskip
\centerline{$AdS_3$ and Linear Dilaton Backgrounds}
}}
\medskip

\centerline{\it A. Giveon${}^{1}$,
D. Kutasov${}^2$, E. Rabinovici${}^{1}$ and A. Sever${}^1$}
\bigskip
\centerline{${}^1$Racah Institute of Physics, The Hebrew University,
Jerusalem, 91904, Israel}

\centerline{${}^2$EFI and Department of Physics, University of Chicago,}

\centerline{5640 S. Ellis Ave., Chicago, IL 60637, USA}

\smallskip

\vglue .3cm
\bigskip

\noindent
We show that string theory in $AdS_3$ has two distinct phases 
depending on the radius of curvature $R_{AdS}=\sqrt{k}l_s$.
For $k>1$ (\ie\ $R_{AdS}>l_s$), the $SL(2,\IC)$ invariant
vacuum of the spacetime conformal field theory is
normalizable, the high energy density of states is given
by the Cardy formula with $c_{\rm eff}=c$, and generic
high energy states look like large BTZ black holes. For
$k<1$, the $SL(2,\IC)$ invariant vacuum as well as BTZ
black holes are non-normalizable, $c_{\rm eff}<c$, and
high energy states correspond to long strings that
extend to the boundary of $AdS_3$ and become more
and more weakly coupled there. A similar picture is
found in asymptotically linear dilaton spacetime with
dilaton gradient $Q=\sqrt{2\over k}$. The entropy
grows linearly with the energy in this case (for $k>\half$).
The states responsible for this growth are two dimensional black
holes for $k>1$, and highly excited perturbative strings
living in the linear dilaton throat for $k<1$. The change
of behavior at $k=1$ in the two cases is an example of a
string/black hole transition. The entropies of black holes
and strings coincide at $k=1$.

\Date{3/05}

\newsec{Introduction}

It is widely believed that in weakly coupled
string theory in flat spacetime, a generic high
energy state with particular values of mass,
angular momentum and other charges looks from
afar like a black hole with those charges. As
the energy of the state increases, the curvature
at the horizon decreases, and the black hole
description becomes more and more reliable.

Conversely, as the mass of a black hole decreases, string
($\alpha'$) corrections in the vicinity of the horizon grow
and the black hole picture becomes less and less appropriate.
When $\alpha'$ effects at the horizon are large, \eg\
if the string frame curvature at the horizon is much larger
than the string scale, a more useful description of the state
is in terms of weakly coupled strings and D-branes. The transition
between the black hole picture valid for small horizon
curvature, and the perturbative string picture valid for large
curvature, which occurs when the curvature at the horizon is
of order the string scale, is smooth. In particular, the black
hole and perturbative string entropies match, up to numerical
coefficients independent of the string coupling and charges.\foot{Both
pictures receive large corrections in the transition region, so
it is difficult to compute and compare these numerical coefficients.}
This is known as the string/black hole correspondence principle
\refs{\SusskindWS,\HorowitzNW}.

The above discussion relies crucially on the fact that
the curvature at the horizon of a black hole depends on
its mass, and goes to zero for large mass. In this paper 
we will study situations where this is not the case.
Two prototypical examples which we will discuss are the
BTZ black hole in $AdS_3$, and the $SL(2,\IR)/U(1)$ black
hole in $\IR^{1,1}$ with asymptotically linear (spacelike)
dilaton. In the former case, the curvature is the same
everywhere, and is proportional to the cosmological constant 
of the underlying anti-de-Sitter spacetime. In the latter, 
the mass of the black hole determines the value of the dilaton 
at the horizon, while the curvature there as well as the gradient
of the dilaton depend solely on the cosmological constant.
Thus, for these black holes varying the mass does not change
the size of the $\alpha'$ corrections.

It is natural to ask whether there is an analog of the string/black
hole transition in these cases. We will see below that the answer
is affirmative. To study the transition, we will take a slightly
different approach than that of flat spacetime. Instead of varying
the energy of the string or black hole, we will study the
dependence of the high energy spectrum on the cosmological constant,
which controls the $\alpha'$ corrections.

For $AdS_3$, this means varying the radius of curvature of the
anti-de-Sitter space, which is related to the cosmological
constant $\Lambda$ via
\eqn\rraaddss{\Lambda=-{1\over R_{AdS}^2}~.}
According to the AdS/CFT correspondence \AharonyTI, the spacetime
dynamics is in this case given by a two dimensional conformal field
theory with central charge \BrownNW
\eqn\cbh{c={3R_{AdS}\over 2l_p}~,}
where $l_p$ is the three dimensional Planck length. The expression
for the central charge, \cbh, is valid when $R_{AdS}$ is much larger
than $l_p$, $l_s$ ($l_s=\sqrt{\alpha'}$ is the string length). We will
usually assume below that the theory is in the semiclassical regime,
$R_{AdS}\gg l_p$, such that the central charge $c$ \cbh\ is very large.
We expect our main results to be independent of that assumption.

The entropy of the two dimensional spacetime CFT behaves at
high energies as
\eqn\sss{S=2\pi\sqrt{c_{\rm eff}L_0\over6}+
2\pi\sqrt{\bar c_{\rm eff}\bar L_0\over6}~,}
where $L_0$ and $\bar L_0$ are the left and right moving
spacetime scaling dimensions. They are related to energy
and angular momentum in $AdS_3$ as follows:
\eqn\enangmom{ER_{AdS}=L_0+\bar L_0-{c\over12};\;\;\;J=L_0-\bar L_0~.}
Modular invariance of the spacetime CFT, which is generally
expected to be a property of quantum gravity in $AdS_3$,
implies that the effective central charge $c_{\rm eff}$
is given by \refs{\KutasovSV,\KutasovPV}
\eqn\ceffff{c_{\rm eff}=c-24\Delta_{\rm min}~,}
where $\Delta_{\rm min}$ is the lowest scaling dimension
in the spacetime theory. Unitarity of the theory implies that
$\Delta_{\rm min}\ge 0$, so that $c_{\rm eff}\le c$; $c_{\rm eff}=c$
iff the $SL(2,\IC)$ invariant vacuum is normalizable, in which case
$\Delta_{\rm min}=0$. Unitary CFT's with $c_{\rm eff}<c$ are known
to exist. Examples include Liouville theory (see \eg\
\refs{\TeschnerRV,\NakayamaVK} for reviews) and the Euclidean CFT
on the cigar, $SL(2,\IR)/U(1)$.

To study string theory in $AdS_3$ with string scale curvatures,
we define the dimensionless ratio
\eqn\rads{\sqrt{k}={R_{AdS}\over l_s}~.}
We will see that the structure of the theory depends crucially
on the magnitude of $k$. For $k>1$, the $SL(2,\IC)$ invariant
vacuum of the spacetime CFT is normalizable; hence the entropy
behaves as \sss\ with $c_{\rm eff}=c$ \ceffff. Generic high
energy states look like BTZ black holes with mass and angular
momentum given by \enangmom. They are non-perturbative
from the point of view of weakly coupled string theory in $AdS_3$.

For $k<1$, the $SL(2,\IC)$ invariant vacuum, as well as BTZ black
holes are non-normalizable; hence, $c_{\rm eff}<c$. The high energy
spectrum is dominated by highly excited states of long strings,
which become more and more weakly interacting as they approach
the boundary. These states are well described by perturbative
string theory in $AdS_3$.

The transition between the black hole and string phases occurs at
the point where the radius of curvature, $R_{AdS}$, is equal to $l_s$
\rads, in agreement with general expectations from the correspondence
principle. At the transition point, the black hole and perturbative
string entropies match exactly.

A similar structure is found for the asymptotically
linear dilaton case. The high energy behavior of the
entropy is in general Hagedorn, but the nature of the 
high energy states depends on the linear dilaton slope, 
$Q$. Below a certain critical
value of $Q$, the generic high energy states correspond to
$SL(2,\IR)/U(1)$ black holes, or their charged cousins.
Above this value, the black holes become non-normalizable,
and the high energy states correspond to highly excited
perturbative strings living in the linear dilaton region.
Again, at the transition between the non-perturbative
(black hole) and perturbative (string) phases, the
entropies match, for both charged and uncharged states.

The fact that the $AdS_3$ and linear dilaton backgrounds
exhibit very similar string/black hole transitions is not
accidental. To see that, consider a background of the form
$\IR^{1,1}\times \IR_\phi\times\CN$, where $\IR_\phi$ is
the real line along which the dilaton varies linearly and 
$\CN$ is a compact unitary CFT.
String theory in this background has in general a Hagedorn
density of states and exhibits the transition described above
as a function of $Q$, the gradient of the dilaton. One can
add to the background $Q_1$ fundamental strings stretched in
$\IR^{1,1}$, and study the low energy dynamics of the combined
system. This theory is related by the AdS/CFT correspondence
to string theory in $AdS_3\times\CN$, with string coupling
$g_s^2\sim 1/Q_1$, and curvature \rads\ $k=2/Q^2$. The
string/black hole transition in $AdS_3\times\CN$ occurs at
$k=1$ for all $Q_1$. Thus, it is natural to expect that the
same transition should also occur at $k=1$ (or $Q=\sqrt2$)
for $Q_1=0$, the original linear dilaton background with
no strings attached. We will see that this is indeed the case.

The $AdS_3$ and linear dilaton theories are also related
via matrix theory \refs{\AharonyTH,\GanorJX}. The low
energy theory on the fundamental strings living in the
linear dilaton throat can be thought of as the discrete
lightcone description of the throat, with the discretization
parameter equal to $Q_1$. This relation implies that
the transitions seen in the two cases should occur
at the same value of $k$.

\newsec{$AdS_3$}

In this section we will study string theory on
$AdS_3$ times a compact space. A large class of
such backgrounds can be constructed as
follows\foot{We will restrict here to a certain
class of supersymmetric backgrounds, but most of
what we say can be generalized to any stable linear
dilaton or $AdS_3$ background.}.
Consider type II string theory on
\eqn\throat{\IR^{1,1}\times\IR_\phi
\times S^1\times \CM~.}
Here, $\IR_\phi$ is the real line labeled by
$\phi$, with a dilaton linear in $\phi$
\eqn\lindil{\Phi=-{Q\over2}\phi~.}
In particular, the string coupling $g_s=\exp\Phi$
goes to zero as $\phi\to\infty$, and diverges when
$\phi\to-\infty$. The worldsheet central charge of
$\phi$ is
\eqn\worldcphi{c_\phi=1+3Q^2~.}
$\CM$ is a compact CFT with $N=2$ worldsheet
supersymmetry and central charge
\eqn\ccmm{c_\CM=9-3Q^2~.}
In the class of backgrounds \throat\ one can perform
a chiral GSO projection which leads to a spacetime
supersymmetric theory, with $N=2$ supersymmetry in
$\IR^{1,1}$ \refs{\KutasovUA,\KutasovPV}.

One can think of \throat\ as the near-horizon geometry
of an $NS5$-brane extended in $\IR^{1,1}$ and wrapped
around a singular four-cycle in a Calabi-Yau fourfold
\GiveonZM. An $AdS_3$ background can be obtained by
adding to it $Q_1$ fundamental strings whose worldvolume
lies in $\IR^{1,1}$. Taking the near-horizon limit of the
strings leads to the geometry
\eqn\adsthree{AdS_3\times S^1\times\CM~.}
The radius of curvature of $AdS_3$, $R_{AdS}$, is related
to the level of the $SL(2,\IR)$ current algebra of $AdS_3$,
$k$, via the relation \rads. The worldsheet central charge
of the CFT on $AdS_3$ is given by\foot{Both here and in
\worldcphi\ we omitted the contribution of the worldsheet
fermions.}
\eqn\adscharge{c_{AdS}=3+{6\over k}~.}
Comparing \throat\ and \adsthree\ we see that
adding the fundamental strings and taking their
near-horizon limit leads to the replacement
\eqn\replaceir{\IR^{1,1}\times\IR_\phi\rightarrow AdS_3~.}
The parameters defining the two models are related via
\eqn\Qk{Q=\sqrt{2\over k}~.}
The relation between the linear dilaton solution
\throat\ and the $AdS_3$ one \adsthree\ is a generalization
of that between the CHS solution describing $NS5$-branes
in flat space \CallanAT\ and the near-horizon geometry of
strings and fivebranes \AharonyTI. Just like there, the growth
of the string coupling as one moves towards the core of the
fivebrane, $\phi\to-\infty$ in \throat, is compensated by the
decrease of the coupling near the fundamental strings. The
combination of the two effects leads to a solution with constant
string coupling
\eqn\ggss{g_s^2\sim {1\over Q_1}~.}
Thus, weakly coupled string theory in $AdS_3$ requires a large
number of fundamental strings, $Q_1\gg1$.

A familiar example of \adsthree\ is the system of $k$
$NS5$-branes wrapped around a four-manifold $\CM_4$
($\CM_4=T^4$ or $K3$), and fundamental strings stretched
in the uncompactified direction of the fivebranes. In this
case one has $\CM={SU(2)_k\over U(1)}\times \CM_4$. The GSO
projection acts as a $\IZ_k$ orbifold on
$S^1\times {SU(2)_k\over U(1)}$, such that the background
\adsthree\ becomes
\eqn\simpads{AdS_3\times S^3\times \CM_4~.}
the levels of $SL(2,\IR)$, \adscharge, and $SU(2)$ are both
equal to the number of fivebranes, $k$, which is an integer
$\ge 2$. Thus, in the class of backgrounds \simpads, the
question what happens for $k<1$ does not arise.

Nevertheless, there are cases where $k<1$ in \adsthree.
For example, we can take $\CM$ to be the $N=2$ minimal
model with $c_\CM=3-{6\over n}$. This corresponds \GiveonZM\
to a singularity of the form $z_1^n+z_2^2+z_3^2+z_4^2+z_5^2=0$
in a Calabi-Yau fourfold, or equivalently to a fivebrane wrapped
around the four dimensional surface $z_1^n+z_2^2+z_3^2=0$, an
$A_{n-1}$ ALE space. Substituting into \ccmm, \adscharge, \Qk,
we find that in this case
\eqn\kknn{k={n\over n+1}<1~.}
Increasing the central charge of $\CM$ beyond $c_{\CM}=3$ takes
one to $k>1$ \GiveonZM.

The spacetime dynamics in an $AdS_3$ background of the
form \adsthree\ corresponds to an $N=2$ superconformal
field theory with spacetime central charge
\refs{\GiveonNS\KutasovXU\GiveonJG-\BerensteinGJ,\GiveonZM}
\eqn\cspacetime{c=6kQ_1~.}
If the string theory in $AdS_3$ is weakly coupled, the
spacetime CFT has a large central charge, proportional to
$1/g_s^2$ \ggss. In this semiclassical regime, the radius
of curvature of $AdS_3$ in Planck units is large,
$R_{AdS}\gg l_p$, \cbh.

The first question we would like to address concerns the
normalizability of the ground state of the spacetime CFT.
{}From perturbative studies of string theory in $AdS_3$ it
is known \KutasovXU\ that the spacetime Virasoro algebra
takes the form
\eqn\spvir{T(x)T(y)\simeq{3kI\over (x-y)^4}+{2T(y)\over(x-y)^2}
+{\partial_yT\over x-y}~,}
where the operator $I$ which determines the spacetime central
charge, $c=6kI$, is given by
\eqn\III{I={1\over k^2}\int d^2z
J(x;z)\bar J(\bar x;\bar z)\Phi_1(x,\bar x;z,\bar z)~,}
in the notations of \KutasovXU. $I$ is an operator which
commutes with the spacetime Virasoro generators. In
particular, it has spacetime scaling dimension zero.
Acting with it on the vacuum of the worldsheet theory
creates a state proportional to the $SL(2,\IC)$ invariant
vacuum of the spacetime theory. Applying \III\ to the
in-vacuum in radial quantization, \ie\ sending $x\to 0$,
we find that the spacetime in-vacuum corresponds to
the worldsheet state
\eqn\stvacuum{
I|0\rangle\sim J^+_{-1}\bar J^+_{-1}|j=0;m=\bar m=-1\rangle~.}
For the out-vacuum (obtained by sending $x\to\infty$)
one gets a similar expression with all $J^3$, $\bar J^3$
charges reversed.

The state \stvacuum\ is a descendant of a principal
discrete series state with $j=0$, $|m|=1$. Such states
are normalizable iff the unitarity condition
\refs{\GiveonPX,\MaldacenaHW}
\eqn\unitcond{-\half<j<{k-1\over2}}
is satisfied. For $k>1$, $j=0$ is in the range \unitcond,
while for $k<1$, it is not. This means that the $SL(2,\IC)$
invariant vacuum of the spacetime CFT is normalizable in the former
case, and is not normalizable in the latter. Hence, due to \ceffff,
for $k>1$ we expect the spacetime theory to have
$c_{\rm eff}=c=6kQ_1$, while for $k<1$ we expect it to have
$c_{\rm eff}<c$. We will calculate $c_{\rm eff}$ for $k<1$
shortly.

It was pointed out in \refs{\StromingerEQ,\BirminghamJT} 
that the Cardy entropy
\sss, with $c_{\rm eff}=c$ given by \cbh, is equal to the
Bekenstein-Hawking entropy of large BTZ black holes, with
mass and angular momentum \enangmom. The above discussion
implies that while for $k>1$ the high energy entropy in the
spacetime CFT corresponding to string theory in $AdS_3$
coincides with the Bekenstein-Hawking one, for $k<1$ the two
are different. Are we to conclude that for $R_{AdS}<l_s$
the string analysis is in disagreement with the
Bekenstein-Hawking prediction?

In fact, it is not. For $k<1$, where $c_{\rm eff}$ and $c$
are different, BTZ black holes are non-normalizable states
and do not belong to the Hilbert space of physical states.
To see that, recall\foot{For recent reviews see \eg\ 
\refs{\BirminghamDT,\CarlipZN}.} that a BTZ black hole 
can be thought of as an orbifold of $AdS_3$ by the discrete 
group $\IZ$ generated by a hyperbolic element of $SL(2,\IR)$,
\eqn\orbtz{g\to h_Lgh_R~,}
with $g\in SL(2,\IR)$,
$h_L=\exp\left[\pi(r_+-r_-)\sigma_3/R_{AdS}\right]$,
$h_R=\exp\left[\pi(r_++r_-)\sigma_3/R_{AdS}\right]$.
Here $r_\pm$ are the inner and outer horizons; they are related
to the mass and angular momentum of the black hole via
\eqn\relhormj{8l_pM={r_+^2+r_-^2\over R_{AdS}^2};\;\;\;
8l_pJ={2r_+r_-\over R_{AdS}}~.}
Since the orbifold \orbtz\ acts on $AdS_3$ as a finite
left and right moving scale transformation, the operator
$I$ \III\ is invariant under \orbtz\ -- it is in the
untwisted sector of the orbifold. Thus, one can repeat
the discussion of the $AdS_3$ vacuum for this case.

The ground state in a sector with a BTZ black hole is again
proportional to $I|0\rangle$, \stvacuum, where now $|0\rangle$
is the worldsheet vacuum of the orbifold theory \orbtz. For
$k>1$, the ground state $I|0\rangle$ is normalizable, while
for $k<1$ it is not, due to \unitcond. Thus, in the latter
case the BTZ black hole is not in the spectrum of the theory.

To recapitulate, we find that string theory in $AdS_3$ has two
phases. In one, corresponding to $R_{AdS}>l_s$ ($k>1$, \rads),
the $SL(2,\IC)$ invariant vacuum is normalizable, the Cardy
formula \sss\ with $c_{\rm eff}=c$ is valid and agrees with the
Bekenstein-Hawking entropy for large BTZ black holes, which are
normalizable as well. In the other, corresponding to $R_{AdS}<l_s$
($k<1$), the $SL(2,\IC)$ invariant vacuum is non-normalizable,
$c_{\rm eff}<c$, but there is no contradiction with the
Bekenstein-Hawking analysis, since BTZ black holes are
non-normalizable in this regime.

It remains to calculate $c_{\rm eff}$ for $k<1$, and to identify
the states in string theory on $AdS_3$ that give rise to the
Cardy entropy \sss\ in this case. This is the problem we turn
to next.

As mentioned in the introduction, one might expect that when
the curvature exceeds the string scale and BTZ black holes
cease to exist as normalizable states, the high energy spectrum
should be dominated by weakly coupled highly excited perturbative
string states. To see whether this is indeed the case, we need to
recall a few facts about the perturbative string spectrum in $AdS_3$.

Low lying states in the background \adsthree\
belong to principal discrete series representations.\foot{Principal
continuous series states correspond to ``bad'' tachyons, which are
absent in the supersymmetric theories discussed here.} The worldsheet
mass-shell condition for these states relates the quadratic Casimir
of $SL(2,\IR)$ to the excitation level $N$ \GiveonNS:
\eqn\wsmass{-{j(j+1)\over k}+N={1\over2}~.}
The scaling dimension of the corresponding Virasoro primary in the
spacetime CFT is
\eqn\hjshort{h=j+1~.}
The unitarity condition \unitcond\ implies that there is an upper
bound on the excitation level $N$ for which \wsmass\ can be solved.
For higher excitation levels and spacetime scaling dimensions, one
needs to consider long string states, which wind around the circle
near the boundary of $AdS_3$ \MaldacenaHW. In a sector with given
winding $w$, the worldsheet mass-shell condition becomes
\eqn\jwlong{-{j(j+1)\over k}-mw-{k\over4}w^2+N={1\over2}~,}
and the spacetime scaling dimension is
\eqn\hwm{h=|m+{k\over2}w|~.}
The quantum number $j$ takes the form $j=-{1\over2}+ip$;
$p\in\IR$ is the momentum of the long string in the
radial direction of $AdS_3$. There are also physical states
of the form \jwlong\ with $j$ real in the range \unitcond,
which correspond to strings that do not make it all the way to
the boundary \MaldacenaHW.

As a first step towards examining the dynamics of long strings,
consider a single string with winding number $w=1$. It can be
thought of as extended in the $\IR^{1,1}$ in \throat.
The remaining, transverse, directions of space,
$\IR_\phi\times S^1\times\CM$, parametrize the target space of the
low energy CFT living on the string. The central charge of that
CFT is $c_l=6k$ \refs{\GiveonNS,\KutasovXU}. The scalar field
$\phi$ parametrizing the location of the string in the radial
direction of $AdS_3$ is described near the boundary by an
asymptotically linear dilaton CFT, with $\Phi=-{Q_l\over2}\phi$
as $\phi\to\infty$. The linear dilaton slope $Q_l$ is given by \SeibergXZ:
\eqn\qqlloo{Q_l=(1-k)\sqrt{2\over k}~.}
Since this result is important for our purposes, we briefly
review its derivation.

The fact that the CFT living on a single long string has
central charge $c_l=6k$ places a constraint on $Q_l$:
\eqn\constql{\left({3\over2}+3Q_l^2\right)+{3\over2}+c_\CM=6k~.}
The term in brackets is the contribution of $\phi$ and its
worldsheet superpartner; the other two contributions on the
l.h.s. are due to $S^1$ and $\CM$, respectively. Using \ccmm,
\Qk\ we have $c_\CM=9-{6\over k}$; substituting in \constql\
we find
\eqn\qlsq{Q_l^2={2\over k}(k-1)^2~.}
In taking the square root of \qlsq, there is a sign ambiguity.
To see that the correct choice is \qqlloo, one can, for example,
compare the dimensions of chiral operators computed in \GiveonZM\
to those computed in the CFT on the long string,
$\IR_\phi\times S^1\times \CM$. This is discussed in appendix A.

The most important features of \qqlloo\ are its overall sign,
and the fact that this sign changes at $k=1$. For $k>1$, the
case discussed in \SeibergXZ, the sign is such that the string
coupling grows when one approaches the boundary of $AdS_3$,
$\phi\to\infty$. This means that highly excited perturbative
strings in $AdS_3$ interact strongly and are unstable near the
boundary (like widely separated quarks and gluons in QCD).

In particular, consider a state with winding number $w$ consisting
of $w$ highly excited long strings near the boundary.
One expects large interactions among the strings to lead to the
formation of a bound state, whose properties can be very different
from those of free strings. A natural candidate for
such a state is a large BTZ black hole, with the same energy and
angular momentum as the original configuration of strings. In this
regime ($k>1$), the perturbative description in terms of highly excited
strings satisfying \jwlong\ is not useful. A better description of the
highly excited state is as a BTZ black hole, which is a non-perturbative
state from the point of view of string theory in $AdS_3$.

For $k<1$, the situation is reversed. The strongly coupled bound state,
the BTZ black hole, is no longer normalizable. Correspondingly, the
interactions among highly excited perturbative strings are now {\it small}
near the boundary, since the gradient of the dilaton on the long strings
\qqlloo\ has the opposite sign. Therefore, the generic high energy state
with winding number $w$ consists of $w$ weakly interacting long strings.
The higher the energy, the smaller the interactions among the strings.
Note that it is crucial for the consistency of the picture that the
linear dilaton on long strings \qqlloo\ change sign precisely
at the same value of $k$ at which the BTZ black hole ceases to be
normalizable. The fact that this is indeed the case supports
the overall picture.

Since for $k<1$ the high energy states consist of weakly interacting
strings, it should be possible to compute the high energy entropy,
and in particular $c_{\rm eff}$ \sss, perturbatively. A quick way to
get the answer is to note that if on a single long string we have the
CFT $\IR_\phi\times S^1\times\CM$, then on $w$ weakly interacting
strings we expect to find the symmetric product CFT
\eqn\symprod{\left(\IR_\phi\times S^1\times\CM\right)^w/S_w~.}
The fact that the interactions among the strings go to zero
as one approaches the boundary of $AdS_3$ should translate in
the spacetime CFT to the statement that \symprod\ describes
accurately the high energy (large $L_0,\bar L_0$) behavior of
the theory of $w$ strings. It may be modified significantly at
low energies.

To calculate the high energy density of states of \symprod\ one
simply adds the effective central charges of the component CFT's
in \symprod: \eqn\ceffwst{c_{\rm eff}=
w\left({3\over2}+{3\over2}+c_\CM\right)= 6w(2-{1\over k})~.}
The winding number $w$ is bounded from above by $Q_1$, the total 
number of strings used to make the background \adsthree. Thus, 
the high energy density of states is
\eqn\cefffull{c_{\rm eff}=6Q_1(2-{1\over k})~.}
This should be compared with the central charge of the
spacetime CFT, $c=6Q_1k$ \cspacetime. Note that one always has
$c_{\rm eff}\le c$, with $c_{\rm eff}=c$ iff $k=1$. This is
required by unitarity and modular invariance of
the spacetime CFT (see the discussion after \ceffff).

 \fig{The behavior of $c$ and $c_{\rm eff}$ as a function of $k$.
 The solid lines represent the dominant contributions to the high
 energy entropy, which are due to BTZ black holes for $k>1$
 (black) and fundamental strings for $k<1$ (red). The dashed lines
 represent states that do not exist in the two phases: black holes
 for $k<1$ (black) and weakly coupled highly excited strings for
 $k>1$ (red).} {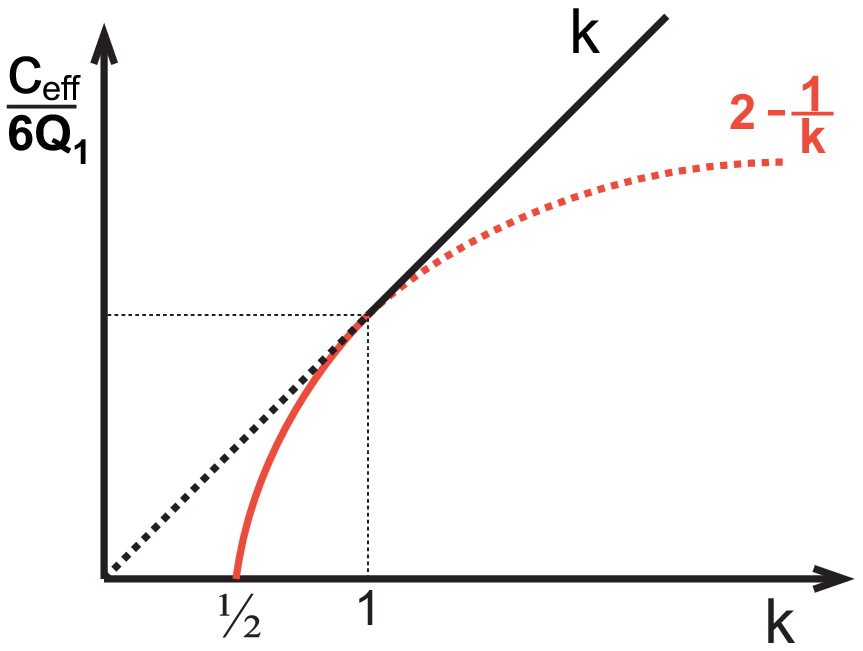}{10 truecm}
 \figlabel\ceff

\noindent
The behavior of the central charge $c$ and $c_{\rm eff}$ as a
function of $k$ is plotted in figure 1. The dashed red line for
$k>1$ represents the contribution of highly excited perturbative
string states in $AdS_3$. As explained earlier, these states are
in fact not there due to the large interactions near the boundary,
and the spectrum in this regime is instead dominated by large
BTZ black holes, whose contribution is represented by the solid
black line in the figure. For $k<1$ the dashed black line corresponds
to the contribution of large BTZ black holes. These are
non-normalizable in this regime; hence, their contribution is
absent. The spectrum is dominated by perturbative strings (solid
red line). Note that in the class of constructions \adsthree, one
is restricted to $k\ge {2\over3}$, since otherwise $\CM$ has a
negative central charge and the theory is non-unitary. In fact, it
is easy to see that in string theory in $AdS_3$ one can never reach
the point $k=\half$, at which $c_{\rm eff}$ vanishes.

Another way to arrive at \ceffwst\ is the following.
We saw before that in the sector with winding $w$, the
worldsheet mass-shell condition takes the form \jwlong.
The spacetime scaling dimension \hwm\ can thus be written
as\foot{We take both $m$ and $w$ to be positive.}
\eqn\hhww{
h={kw\over4}+{1\over w}\left[-{j(j+1)\over k}+N-{1\over2}\right]~.}
The bulk of the high energy density of states comes from the
entropy associated with the excitation level, $N$. To estimate it,
we can assume that the term proportional to $N$ dominates the
r.h.s. of \hhww, so that
\eqn\hwNN{h\simeq{N\over w}~.}
The density of states grows with $N$ in a way familiar
from perturbative string theory \KutasovSV. The effective
worldsheet central charge of \adsthree\ is
$c_{\rm eff}^{({\rm ws})}=6+c_\CM-3$; the $6$ comes from
$AdS_3\times S^1$, while the $-3$ accounts for the two
towers of bosonic and fermionic oscillators removed by
gauge invariance. Substituting \ccmm\ we find that
$c_{\rm eff}^{({\rm ws})}=6(2-{1\over k})$. Using the
fact that the spacetime scaling dimension $h$ is related
to $N$ via \hwNN, we see that the perturbative string
entropy, $2\pi\sqrt{c_{\rm eff}^{({\rm ws})}N/6}$,
gives rise to a spacetime entropy \sss\ with
$c_{\rm eff}=wc_{\rm eff}^{({\rm ws})}$,
in agreement with \ceffwst.

The spectrum of long strings gives the same effective central charge
as \symprod, but does the spacetime CFT approach the symmetric product
as $L_0,\bar L_0\to\infty$? We will leave a detailed analysis of this
question to future work.\foot{For a discussion of the spectrum in the
$w\not=0$ sectors and the relation to symmetric orbifolds, see \ArgurioTB.}
Here, we note that
the dimension formula \hhww\ can be written in the following suggestive
way. Consider a physical state in the sector with winding number one,
which satisfies \hhww\ with $w=1$. Denote its spacetime scaling dimension
by $h_1$. According to \hhww, for every such state there is a state in the
sector with winding $w$, related to it by spectral flow, whose dimension
$h_w$ is given by:
\eqn\hwhone{h_w={h_1\over w}+{k\over4}\left(w-{1\over w}\right).}
This is precisely the type of expression one finds
in the $w$-twisted sector of a symmetric orbifold, 
whose building block has $c=6k$ \refs{\KlemmDF,\FuchsVU,\ArgurioTB}.

We see that a number of seemingly independent things happen at $k=1$:
\item{(1)} The BTZ black holes, which are normalizable for $k>1$,
cease to be so for $k<1$. The same is true for the $SL(2,\IC)$
invariant vacuum of the spacetime theory.
\item{(2)} The gradient of the linear dilaton on a long string
living near the boundary of $AdS_3$ \qqlloo\ changes sign at $k=1$.
For $k>1$, long strings are strongly coupled near the boundary,
while for $k<1$ they are weakly coupled there.
\item{(3)} The black hole entropy, which is given by the Cardy
formula \sss\ with $c_{\rm eff}=c=6Q_1k$ \cspacetime, coincides
(see figure 1) with the perturbative string entropy, which has
$c_{\rm eff}=6Q_1(2-{1\over k})$ \cefffull.

\noindent
We argued above that these phenomena are in fact related. They
are all manifestations of a sharp transition, which occurs at $k=1$,
between a strongly coupled phase where the high energy states
are large black holes, and a weakly coupled phase in which they are
weakly interacting fundamental strings. The matching of the
entropies suggests that the dynamics of large BTZ black holes
changes smoothly into that of highly excited fundamental strings
as we vary $k$.

We finish this section with a few comments. So far, we
discussed type II string theory in $AdS_3$. One could ask
whether the matching that we found of black hole and string
entropies extends to other cases, such as the heterotic string.
Heterotic strings in $AdS_3$ were studied in \KutasovZH, where
it was shown that in this case the spacetime CFT inherits the
chiral nature of the worldsheet one. While the left
moving\foot{We take the left-movers to be supersymmetric and
the right-movers bosonic.} central charge is still given by
\cspacetime, $c=6Q_1k$, the right moving one is equal to
$\bar c=6Q_1(k+2)$. In the black hole phase this means that the
Cardy formula is \sss, with $c_{\rm eff}=c$ and $\bar c_{\rm eff}=\bar c$.

In the perturbative string phase, the left-movers still satisfy
\jwlong, \hwm, \hhww, while for the right-movers one has the
mass-shell condition
\eqn\bosmass{-{j(j+1)\over k}-\bar mw-{k+2\over4}w^2+\bar N=1~.}
For large excitation level $\bar N$ one has, as in
\hwNN, $\bar h\simeq \bar N/w$.
To calculate the effective worldsheet central charge of the
right-movers recall that they are described by the worldsheet
CFT
\eqn\rightcft{AdS_3\times S^1\times \CM\times\Gamma~,}
where $\Gamma$ is a CFT with central charge 13, which
completes the total central charge of \rightcft\ to 26:
\eqn\cbalance{3+{6\over k}+1+c_\CM+13=26~.}
The corresponding effective worldsheet central charge is
$\bar c_{\rm eff}^{({\rm ws})}=3+1+c_\CM+13-2=24-{6\over k}$.
Multiplying by $Q_1$ we conclude that the analog of \cefffull\
for the right-movers in the heterotic string is
\eqn\barceff{\bar c_{\rm eff}=6Q_1(4-{1\over k})~.}
We see that, like in the type II case, $\bar c_{\rm eff}$
is always smaller than or equal to $\bar c=6Q_1(k+2)$; the two
coincide at the string/black hole transition point $k=1$, in
agreement with the general picture presented above.

Another natural question concerns the relation \ceffff\ for $k<1$.
In that case, we found $c=6Q_1k$ \cspacetime, and
$c_{\rm eff}=6Q_1(2-{1\over k})$ \cefffull. According to
\ceffff, this means that
\eqn\mindim{\Delta_{\rm min}={Q_1\over4}{(k-1)^2\over k}=
Q_1{Q_l^2\over8}~.}
This is puzzling since, especially for large $Q_1$, there are
many normalizable states of the form \jwlong, \hwm, whose
dimensions are smaller than \mindim.

One possible resolution of this conundrum is the following. It
was pointed out in \GiveonUP\ that in perturbative string theory
in $AdS_3$, the spacetime central charge is in fact not proportional
to the identity operator, but takes different values in different
sectors of the theory. In other words, the spacetime CFT appears
to have a non-trivial vacuum structure.

The situation is analogous to studying the extreme IR limit of an 
$N=2$ Landau-Ginzburg model of a chiral superfield $\Phi$ with 
superpotential 
\eqn\wphi{W=\sum_{i=1}^n a_i\Phi^i~.}
For generic values of the $\{a_i\}$, the IR limit is a direct sum
of massive theories. By fine tuning the coefficients one can reach
multicritical points for which 
\eqn\wphimulti{W=a_n\prod_j(\Phi-\Phi_j)^{n_j}~,\qquad n_j\ge2~,
\qquad \sum_j n_j=n~,}
and the theory is a direct sum of minimal models 
\eqn\dirsum{\oplus_j\CM_{n_j}~.}
Thus, it exhibits non-trivial vacuum structure. In 
particular, the torus partition sum is given by a 
sum of the partition sums of the individual $N=2$
minimal models, $\CZ=\sum_j \CZ_{n_j}$.
 
In our case, it is possible that the spacetime CFT for 
$k<1$ splits into sectors labeled by winding numbers 
$(w_1,w_2,\cdots)$, with $\sum_jw_j=Q_1$. In a given sector,
the spacetime CFT is a direct sum of the individual theories
corresponding to the different $w_j$. Each of those has 
central charge $c^{(w_j)}(=6w_jk)$, and 
$c_{\rm eff}^{(w_j)}(=6w_j(2-{1\over k}))$.
The torus partition sum is given by a sum over different
components,
\eqn\ztau{\CZ(\tau,\bar\tau)=\sum_j\CZ_{w_j}(\tau,\bar\tau)~.}
The partition sums $\CZ_{w_j}$ are separately modular
invariant. Thus, instead of a single minimal dimension 
operator \mindim, we have a separate minimal dimension
operator for each $w$, with
\eqn\mindimww{\Delta_{\rm min}^{(w)}={1\over24}
\left( c^{(w)}-c_{\rm eff}^{(w)}\right)=
{w\over4}{(k-1)^2\over k}=
w{Q_l^2\over8}~.}
Clearly, more work is required to see whether this is the right
picture. As a first step, in appendix B we show that perturbative 
string states in the sector with winding $w$ indeed have spacetime
scaling dimensions which are bigger than $\Delta_{\rm min}^{(w)}$
\mindimww.

In this section we studied string theory in $AdS_3$ with only
the $NS$ $B_{\mu\nu}$ field turned on. This is usually viewed
as a special locus in the moduli space of string theory in
$AdS_3$, along which the spacetime CFT is singular \SeibergXZ.
The singular nature of the spacetime CFT is reflected in the
appearance of a continuum of scaling dimensions above a gap,
associated with long strings. To eliminate the continuum, one
can often turn on a modulus that creates a potential for long
strings, preventing them from exploring the region near the
boundary, where they are strongly coupled.

For example, for the background $AdS_3\times S^3\times\CM_4$
\simpads, the modulus in question is the $N=4$ Liouville
superpotential on the long strings, which grows as one approaches
the boundary of $AdS_3$, and thus repels the strings from this
region \SeibergXZ. This modulus is described by a RR vertex
operator in perturbative string theory in $AdS_3$ \GiveonNS.

It is important to emphasize that the above picture is only valid for $k>1$.
In the weakly coupled phase $k<1$, similar RR perturbations grow as one
moves {\it away} from the boundary of $AdS_3$, and thus do not change the
singular nature of the spacetime CFT. In essence, for $k<1$ the spacetime
CFT is {\it always} singular, \ie\ it always has a continuum above a gap
in its spectrum of scaling dimensions.

Note that this behavior is consistent with the general picture
described above: for $k>1$, perturbative strings in $AdS_3$ are strongly
coupled at high energies, and the strength of their interactions can be
modified by changing the moduli of the spacetime CFT. For $k<1$, perturbative
strings are weakly coupled at high energies and thus no marginal or relevant
perturbations can make their interactions strong there.

\newsec{Linear dilaton}

In the previous section we have seen that the low
energy physics in linear dilaton backgrounds such
as \throat, in the presence of $Q_1$ fundamental
strings in the throat, is qualitatively different
for $k<1$ and $k>1$. This was demonstrated for
$Q_1\gg1$, but we expect it to be true for all
$Q_1$. It is natural to ask whether something similar
happens for $Q_1=0$, \ie\ in the absence of strings in
the throat. In this case, the theory under consideration
is the Little String Theory (LST) of the throat
\refs{\AharonyUB,\GiveonZM}, and one would like to know
whether the nature of the states which dominate its high
energy behavior changes as $k$ passes through the value
$k=1$, or as the gradient of the dilaton $Q$ passes
through $Q=\sqrt2$. We will next show that this is indeed
the case.

The high energy density of states in LST exhibits
Hagedorn growth (see \eg\ \refs{\KutasovJP,\AharonyXN}
for recent discussions). For $k>1$, this comes from the
Bekenstein-Hawking density of states of $SL(2,\IR)_k/U(1)$
black holes. These are obtained by considering the string
frame dilaton gravity action
\eqn\dilgr{{\cal S}=\int d^2x \sqrt{-g}e^{-2\Phi}
\left(\RR+4g^{\mu\nu}\partial_\mu \Phi\partial_\nu
\Phi-4\Lambda\right)~,}
where, as before, $\Lambda$ is related to the level of
$SL(2,\IR)$, $k$, by \rraaddss, \rads. The equations
of motion of \dilgr\ have the solution\foot{Here and 
below we often set $\alpha'=2$.} \refs{\WittenYR,\MandalTZ}
\eqn\sltwobh{\eqalign{
&ds^2=f^{-1}d\phi^2-fdt^2~, \qquad f=1-{2M\over r}~,\cr
&Qe^{-2\Phi}=Qe^{Q\phi}=r~.\cr
}}
Here $(t,\phi)$ provide a parametrization of the two
dimensional space $\IR_t\times\IR_\phi$ in \throat;
$(t,r)$ are analogs of Schwarzschild coordinates.
The value of the dilaton at the horizon of the black
hole, $\Phi(r=2M)=\Phi_0$, is determined by its mass
$M$: $\exp(-2\Phi_0)=2M/Q$. The gradient of the dilaton
$Q$ is related to the level $k$ via \Qk. Note that, as
mentioned in the introduction, the mass determines the
strength of quantum corrections near the horizon, but
the $\alpha'$ effects are insensitive to it.

The Bekenstein-Hawking entropy of these black holes is
obtained in the usual way by Wick rotating \sltwobh\ to
Euclidean space and calculating the circumference of the
Euclidean time direction near the boundary at $\phi=\infty$.
After Wick rotation, the metric \sltwobh\ can be written as
\refs{\ElitzurCB,\WittenYR}
\eqn\euclideanbh{\eqalign{
ds^2=&kl_s^2\left(d\rho^2+\tanh^2 \rho d\tau^2\right)\cr
\Phi=&\Phi_0-\log\cosh\rho~,\cr
}}
where $\tau$ is periodic, with period $2\pi$.
The corresponding entropy is
\eqn\bhentropy{S_{\rm bh}=2\pi l_sM\sqrt{k}~.}

\noindent
{}From the string/black hole correspondence principle and
the discussion of the previous section one would expect
that when $\alpha'$ corrections to the geometry \sltwobh,
\euclideanbh\ become of order one, the black hole should
become non-normalizable and the system should make a
transition to a string phase.

This leads to the question what is a useful measure
of the $\alpha'$ corrections. The simplest guess is
the curvature at the horizon \refs{\SusskindWS,\HorowitzNW}.
In two dimensions, every spacetime is maximally symmetric.
Thus, the Riemann tensor is completely determined by the
scalar curvature, and one may try to check when the latter
becomes of order the string scale. For \sltwobh, the scalar
curvature is
\eqn\curva{\RR(r)={2\over k}{2M\over r}~.}
At the horizon, $\RR (r=2M)={2\over k}$, and one may expect
a transition to a string phase at $k\sim 1$. In fact, it
turns out that the curvature is not the most useful guide for
estimating the size of string corrections. One way to see
that is to consider the generalization of \curva\ to charged
black holes. We will see below that in that case, as one
varies the mass to charge ratio, the curvature at the outer
horizon changes continuously and even vanishes for a
particular ratio. This does not necessarily mean that string
corrections are small there!

Indeed, the metric is not the only field that is excited in the
black hole geometry. In the uncharged case \sltwobh\ we also
have a non-trivial dilaton, and for charged black holes a gauge
field as well. We are looking for a measure of the size of the
$\alpha'$ corrections which takes all these fields into account.

It is natural to propose that the relevant quantity is the
cosmological constant $\Lambda$ in \dilgr. This parameter
determines both the gradient of the dilaton, and the curvature
of the metric, and it plays the same role here as in the
$AdS_3$ discussion of section 2. Furthermore, while the
curvature \curva\ depends on position, the cosmological
constant does not, so we do not have to face the issue
where in the black hole spacetime to evaluate it.

A nice feature of the above proposal is that the black hole
\sltwobh\ is a coset of $SL(2,\IR)$, while the BTZ black hole
discussed in section 2 is an orbifold of the same space.
Therefore, it is natural to expect that the $\alpha'$ corrections
should be determined by the underlying $SL(2,\IR)$ group manifold,
and should therefore only depend on $k$, \rraaddss, \rads, \Qk.
It also makes it easy to generalize to other examples related to
$SL(2,\IR)$, such as the charged black holes that will be
discussed below.

Another question that one might ask at this point is the following.
While $\alpha'$ corrections are expected to grow when $Q$ increases
(or $k$ decreases), it is known \bars\ that the metric and dilaton
\sltwobh, \euclideanbh, which are obtained by studying the lowest
order in $\alpha'$ action \dilgr, give rise to a solution of the
classical string equations of motion\foot{This is the case for
fermionic strings; in the bosonic case, there are corrections, which
were discussed in \DijkgraafBA.}, to all orders in $Q$. In what sense
are $\alpha'$ corrections growing\foot{This question could
have been asked in $AdS_3$ as well, but we chose to discuss
it here, since it is more familiar in this case.} as one increases $Q$?

The answer to this question is that for large $Q$, or small $k$,
the sigma model description breaks down due to non-perturbative effects
in $Q$ or $1/k$. For example, to describe the Euclidean black hole
\euclideanbh\ in the bosonic string, one needs to turn on, in
addition to the metric and dilaton, certain winding modes of the
tachyon. This was first shown in unpublished work by V. Fateev, A.
Zamolodchikov and Al. Zamolodchikov, and is reviewed in \KazakovPM;
see also \KarczmarekBW\ for a recent discussion. In the fermionic
string, which is the case relevant here, one has to turn on winding
modes of the fermionic string tachyon \GiveonPX. The leading one is
the worldsheet superpotential
\eqn\wssuperpot{\int d^2\vartheta e^{-{1\over Q}\Psi}~,}
where $\Psi$ is a chiral superfield whose bottom component is
\eqn\lowcomppsi{\Psi=\phi+i\sqrt{k}(\tau_L-\tau_R)+\cdots.}
The worldsheet superpotential \wssuperpot\ is non-perturbative
in $Q$ and becomes more and more important as $Q$ grows.
Its effects are visible in the structure of correlation functions
in the Euclidean black hole geometry, as discussed in \GiveonUP.

An interesting question is what is the analog of the superpotential
\wssuperpot\ for the Minkowski $SL(2,\IR)\over U(1)$ black hole
\sltwobh. This question is rather confusing, since the mode
\wssuperpot\ has winding number one around the Euclidean time
direction, and is difficult to continue to Minkowski spacetime.

One might be able to make sense of this continuation by using the
observation that T-duality exchanges the region outside the
horizon of the Minkowski $SL(2,\IR)\over U(1)$ black hole with
the region behind the singularity~\refs{\GiveonSY,\DijkgraafBA}.
One can try to think of the analytic continuation of the winding
mode \wssuperpot\ as a finite energy excitation living ``behind
the singularity at $r=0$'' in the extended black hole spacetime.
An observer living far outside the horizon of such a black hole
would not see this field turned on directly, but would presumably
feel its effects on the physics.

Now that we have established that as $k$ decreases, $\alpha'$ effects
grow, and pointed out the close analogy to the $SL(2,\IR)$ case of
section 2, we can turn to a discussion of the string/black hole
transition. The most important effect of the superpotential \wssuperpot\
is that it implies that for $k<1$, the Euclidean $SL(2,\IR)\over U(1)$ 
black hole is non-normalizable. Indeed,  the perturbation \wssuperpot\ 
is normalizable for ${1\over Q}>{Q\over2}$, and non-normalizable otherwise.
The transition point occurs at $Q^2=2$, or $k=1$.

Thus, for $k<1$ the Euclidean black hole \euclideanbh\ does not
contribute to the canonical partition sum. The behavior \bhentropy,
which is due to its contribution, is only valid for $k>1$.

The discussion of the previous section suggests that for $k<1$
the high energy spectrum should be dominated by highly excited
perturbative strings living in the linear dilaton throat \throat.
The same calculation as that done there gives the entropy of these
strings,
\eqn\psentropy{S_{\rm pert}=2\pi l_sM\sqrt{2-{1\over k}}~.}
Equations \bhentropy\ and \psentropy\ are direct analogs of the
discussion of section 2. The former is the analog of the BTZ black
hole entropy \sss\ with $c_{\rm eff}=6kQ_1$; the latter, of the
perturbative string entropy with \cefffull\
$c_{\rm eff}=6(2-{1\over k})Q_1$. The same comments as there apply
here. In particular, $S_{\rm bh}\ge S_{\rm pert}$; the two are equal
at the string/black hole transition point $k=1$. As in $AdS_3$, the
transition from black holes to strings at $k=1$ occurs not because
the string entropy becomes larger than the black hole one, but rather
because the black holes cease to be normalizable.

A natural question is why the highly excited fundamental strings change
their behavior at $k=1$. In analogy to the $AdS_3$ discussion, we expect
that for $k>1$, as the energy of perturbative strings grows, they should
become more strongly coupled and form bound states, the $SL(2,\IR)\over U(1)$
black holes. For $k<1$ their coupling should go to zero as their energy
goes to infinity. In $AdS_3$ this was a consequence of the dynamics on long
strings, and in particular of the behavior of the linear dilaton slope
\qqlloo.

We will not analyze the analogous question in our case, except to
say that the behavior should be the same here. Note that the central
charge of the CFT on a long string in $AdS_3$, $c=6k$, and the value
of $Q_l$ \qqlloo, are independent of the number of strings that make
the $AdS_3$ background, $Q_1$. As $Q_1$ decreases, the string coupling
in $AdS_3$ grows, but the fact that, for $k<1$, long strings become free
in the UV (\ie\ near the boundary) does not change, since it is driven
by the sign of \qqlloo. Hence, one expects that these results can be
extrapolated to small $Q_1$, and in particular to $Q_1=0$. It would be
nice to show this directly.

To recapitulate, we see that like in $AdS_3$, string theory
in linear dilaton backgrounds, such as \throat, has two distinct phases.
For $k>1$, the high energy entropy is given by \bhentropy\ and is due to
$SL(2,\IR)\over U(1)$ black holes. For $k<1$ these black holes become
non-normalizable, and the entropy, while still Hagedorn (for $k>\half$),
is now given by \psentropy\ and is due to highly excited weakly coupled
fundamental string states living in the linear dilaton throat.

The detailed formulae for the entropy \bhentropy, \psentropy\ bear
a striking resemblance to those found in the $AdS_3$ case. From the
worldsheet point of view, this is due to the fact that in both cases
there is an underlying $SL(2,\IR)$ structure. From the spacetime point
of view one can understand the agreement in the following way. One of
the ways to study LST's corresponding to linear dilaton throats is
using Matrix theory \refs{\AharonyTH,\GanorJX}. This is obtained
by studying the low energy dynamics of $Q_1$ fundamental strings in
the linear dilaton throat, which is exactly the theory discussed in
section 2. $Q_1$ is interpreted as the discretized light-like momentum,
$P^+$, while $P^-$  corresponds to the energy in the CFT describing
the strings in the throat. For uncharged states it is given by
$L_0-{c\over24}=\bar L_0-{c\over24}$. Thus, the energy in LST is 
given by
\eqn\dlcq{{\alpha'\over4}M^2=Q_1\left(L_0-{c\over24}\right)~.}
This provides a direct map between the CFT entropy
\sss\ and the LST entropy \bhentropy, \psentropy, using the
formulae for $c_{\rm eff}$, \cspacetime, \cefffull. In particular,
when the CFT of the strings in the throat makes the transition
from the BTZ black hole phase to the perturbative string phase,
the LST makes a similar transition from the $SL(2,\IR)/U(1)$
black hole phase to the perturbative string one.

In the perturbative string phase, the DLCQ prescription
\dlcq\ with discretization parameter $w$,  and 
the $AdS_3$ result \hhww, give rise to the spectrum
\eqn\llwwqq{{\alpha'\over4} M^2=
w\left[L_0^{(w)}-{c^{(w)}\over24}\right]=
-{j(j+1)\over k}+N-\half~.}
It is interesting that \llwwqq\ is the same as the spectrum of
perturbative strings living in the linear dilaton throat, with
the usual relation between the momentum in the $\phi$ direction
and $j$, $\beta=Qj$ (see appendix A). 

Thus, we see that for $k<1$, DLCQ maps weakly coupled highly
excited fundamental strings in  $AdS_3$ to weakly coupled 
strings in the linear dilaton throat. Similarly, for $k>1$,
when the strings are strongly coupled, DLCQ maps BTZ black holes
in $AdS_3$ to $SL(2,\IR)\over U(1)$ black holes in LST.

So far we discussed black holes that do not carry any charges. As 
a further check on the general picture, we will next show that the
string/black hole transition at $k=1$ works correctly also for a 
certain class of black holes and strings that are charged under one 
or two gauge fields.

\subsec{One charge black holes}

In this subsection we will test the string/black hole correspondence
for two dimensional black holes that carry charge $q$ under an Abelian
gauge field $A_\mu$. The black holes in question are solutions of two
dimensional dilaton gravity \dilgr\ coupled to the gauge field,
\eqn\dilgrav{\CS=\int d^2x\sqrt{-g}e^{-2\Phi}
\left(\RR+4g^{\mu\nu}\partial_\mu
\Phi\partial_\nu\Phi-{1\over 4}F^2-4\Lambda\right)~,}
where $F_{\mu\nu}$ is the field strength of $A_\mu$. The charged
black hole solution that we will consider  is \McGuiganQP\
\eqn\cbhs{ds^2=f^{-1}d\phi^2-fdt^2~, \qquad
Qe^{-2\Phi}=Qe^{Q\phi}=r~,
\qquad A={q\over r}\sqrt2 dt~,}
where
\eqn\fff{f=1-{2M\over r}+{q^2\over r^2}~.}
The values of the dilaton at the inner and outer horizons of the
black hole are related to the mass $M$ and charge $q$ of the black
hole as follows:
\eqn\mmqq{r_{\pm}\equiv Qe^{-2\Phi_\pm}=M\pm\sqrt{M^2-q^2}~.}
For $q=0$, the solution \cbhs\ -- \mmqq\ reduces to the uncharged
one \sltwobh.

The curvature of the metric \cbhs\ is
\eqn\curcha{\RR(r)={2\over k}\left({2M\over r}-{4q^2\over r^2}\right)~.}
In particular, its value at the event horizon,
\eqn\curhor{\RR(r_+)={2\over k}{2Mr_+-4q^2\over 2Mr_+-q^2}~,}
changes continuously from $2/k$, for $q=0$, to $-4/k$ in the
extremal case $M=q$. As mentioned above, this is an example
where the curvature {\it at} the horizon is not a particularly
good guide to the size of $\alpha'$ corrections.
One can still use the curvature for that purpose, by studying
the geometry in the vicinity of $r=r_+$, but as we argued
before, a simpler indicator of the size of string
corrections is the cosmological constant $\Lambda$, which is
sensitive to all the fields via the dilaton equation of motion,
\eqn\lrpf{
\RR -4(\nabla_\mu\Phi\nabla^\mu\Phi-\nabla_\mu\nabla^\mu\Phi)-{1\over 4}F^2
=4\Lambda=-{2\over k}~.}
The string/black hole transition is expected to occur at $k=1$,
when the cosmological constant is equal to the string scale.

To realize the black hole solution \cbhs\ in string theory,
consider a background of the form \throat, and compactify
the spatial direction of $\IR^{1,1}$, replacing it by
$\IR_t\times S^1$. The two dimensional spacetime of \dilgrav\
is $\IR_t\times\IR_\phi$, and the $S^1$ gives rise to two gauge
fields, whose charges are the left and right-moving momenta
(or momentum and winding) on the circle. Here, we are interested
in black holes carrying only one charge, which we will choose to
be the right-moving momentum, $p_R$. The resulting black hole
background \cbhs\ can be described by a coset CFT, like the
uncharged black hole \sltwobh. The coset in question is
\refs{\JohnsonJW,\GiveonGE}
${SL(2,\IR)_k\times U(1)_R\over U(1)}$, where $U(1)_R$
is the right-moving part of the $S^1$ mentioned above.
The $U(1)$ symmetry that is being gauged acts on $SL(2,\IR)\times
U(1)_R$ as
\eqn\uacts{(g,x_R)\simeq (h_Lgh_R,x_R+\beta)~,}
where $(g,x_R)\in SL(2,\IR)\times U(1)_R$ and
\eqn\hlhrxr{
h_L=e^{\alpha\sigma_3}~,\,\,\, h_R=e^{\alpha\cos(\psi)\sigma_3}~,\,\,\,
\beta=\alpha\sin(\psi)~,}
with $\tan^2(\psi/2)=r_-/r_+$.

The entropy of the black hole is given
in terms of its mass and charge by \NappiAS\
\eqn\sbh{S_{\rm bh}=\pi\sqrt{2k}\left(M+\sqrt{M^2-q^2}\right)~.}
The description in terms of a coset of $SL(2,\IR)$ suggests
that the black hole exists as a normalizable state only for
$k>1$, while for $k<1$ the generic high energy states with the same
charge correspond to highly excited perturbative strings with
$(p_L,p_R)=(0,q)$. The entropy of such strings is
\eqn\sst{S_{\rm pert}=
2\pi \sqrt{2-{1\over k}}\left(\sqrt{N}+\sqrt{\bar N}\right)~;}
the left and right moving excitation numbers $(N,\bar N)$ are given 
in terms of $M$ and $q$ by the on-shell condition\foot{Up to additive
corrections of order one.}
\eqn\nnnn{M^2=2N=2\bar N+q^2~.}
Substituting \nnnn\ in \sst\ gives
\eqn\ssst{S_{\rm pert}=\pi\sqrt{4-{2\over k}}\left(M+\sqrt{M^2-q^2}\right)~.}
Comparing \ssst\ to \sbh\ we see that the situation is very similar
to the uncharged case. The black hole entropy, $S_{\rm bh}$, is always
larger or equal than the perturbative string one, $S_{\rm pert}$. The
two coincide at the transition point $k=1$, below which the black hole
does not exist and perturative strings dominate. The uncharged case is
reproduced for the special case $q=0$.

\subsec{Two charge black holes}

In the previous subsection we studied states with only right-moving
charge $p_R$, but no left-moving one. It is natural to generalize
the discussion to states carrying both left and right-moving charges.
On the perturbative string side, these are states with non-zero left
and right-moving momenta $(q_L,q_R)$. Their high energy entropy is given by
\eqn\sqlqr{S_{\rm pert}=\pi\sqrt{4-{2\over k}}\left(\sqrt{M^2-q_L^2}
+\sqrt{M^2-q_R^2}\right)~.}
The corresponding black holes can be described as follows.
Start with three dimensional dilaton gravity coupled to the
Neveu-Schwarz $B$-field,
\eqn\dilgh{
\CS=\int d^3X\sqrt{-G}e^{-2\Phi}
\left(\RR+4G^{MN}\partial_M
\Phi\partial_N\Phi-{1\over 12}H^2-4\Lambda\right)~,}
where the three form $H=dB$ is the field strength of $B$.
This action has a black string solution,
\eqn\bssol{\eqalign{
 G_{xx}=&1+{2q_Rq_L\over Mr}~, \qquad Qe^{-2\Phi}=r~,\cr
 G_{tt}=&-1+{M^2+\sqrt{(M^2-q_R^2)(M^2-q_L^2)}-q_Rq_L\over Mr}~,\cr
 G_{rr}=&\[\(Mr-M^2+q_Rq_L\)^2-(M^2-q_R^2)(M^2-q_L^2)\]^{-1}
         {kM^2\over 2}~,\cr
 G_{tx}=&-{q_L\sqrt{M^2-q_R^2}+q_R\sqrt{M^2-q_L^2}
          \over\[2\sqrt{(M^2-q_R^2)(M^2-q_L^2)}
 \(M^2+\sqrt{(M^2-q_R^2)(M^2-q_L^2)}-q_Rq_L\)\]^{1\over 2}}G_{tt}~,\cr
 B_{tx}=&{q_L\sqrt{M^2-q_R^2}-q_R\sqrt{M^2-q_L^2}\over
         \[2\sqrt{(M^2-q_R^2)(M^2-q_L^2)}
 \(M^2+\sqrt{(M^2-q_R^2)(M^2-q_L^2)}-q_Rq_L\)\]^{1\over 2}}G_{tt}~.}}
Here, $x$ is taken to be compact, $M$ is the ADM mass, 
and $(q_L,q_R)$ are related to the angular momentum, 
$q_G$, and the axion charge per unit length, $q_B$, by
\eqn\qgqb{q_G=q_L+q_R~, \qquad q_B=q_L-q_R~.}
The black string has a singularity hidden behind inner and outer horizons,
all of which are extended in $x$. The singularity is located at $r=0$
and the horizons are at
\eqn\bsrph{r_{\pm}=M\pm{\sqrt{M^2-q_L^2}\sqrt{M^2-q_R^2}-q_Lq_R\over M}~.}
Performing a Kaluza-Klein reduction down to two dimensions $(t,r)$
gives a two dimensional black hole with two charges \qgqb.
The one charge solution of the previous subsection corresponds to
$q_R=q$, $q_L=0$. The static black string solution,
\bssol\ with $q_L+q_R=0$, was studied in \HorneGN.

The black string background \bssol\ can be described as
the quotient CFT ${SL(2,\IR)\times U(1)\over U(1)}$.
The $U(1)$ symmetry that is being gauged acts on
$SL(2,\IR)\times U(1)_L\times U(1)_R$ as follows
\eqn\ugxlxr{(g,x_L,x_R)\simeq (h_Lgh_R,x_L+\beta_L,x_R+\beta_R)~,}
where $(g,x_L,x_R)\in  SL(2,\IR)\times U(1)_L\times U(1)_R$ and
\eqn\hlhrklkr{h_L=e^{\alpha\sqrt{M^2-q_L^2}\sigma_3}~,\qquad
              h_R=e^{\alpha\sqrt{M^2-q_R^2}\sigma_3}~,\qquad
              (\beta_L,\beta_R)=\alpha (q_L,q_R)~.}
Some further discussion of this coset CFT, the black string
solution \bssol, and the relation between them, appear in
appendix C, where it is shown that the Bekenstein-Hawking
entropy of the black string is given by
\eqn\sbs{S_{\rm bs}=
\pi\sqrt{2k}\left(\sqrt{M^2-q_L^2}+\sqrt{M^2-q_R^2}\right)~.}
This is a natural generalization of the uncharged \bhentropy,
and one charge \sbh\ entropies discussed above. Comparing to
\sqlqr\ we see that the relation between the black hole and
string entropies, as well as the matching between them at $k=1$
work in the same way as before.

\newsec{Discussion}

In this paper we argued that string theory in asymptotically
$AdS_3$ and linear dilaton spacetimes exhibits an interesting
phase structure as one varies the (negative) cosmological
constant $\Lambda$ that produces the background. For
$|\Lambda|<|\Lambda_c|$, the theory is in a strongly coupled
phase. Generic high energy states in this phase correspond to
large black holes; they are non-perturbative from the weakly
coupled string perspective. For $|\Lambda|>|\Lambda_c|$, the
black holes become non-normalizable and the generic high energy
states are highly excited weakly coupled perturbative string
states. The critical cosmological constant $\Lambda_c$ is of
order the string scale.

The distinction between the two phases is very reminiscent of the
string/black hole correspondence of \refs{\SusskindWS,\HorowitzNW}.
The main difference is that in that case, the transition between
the black hole and string pictures can be achieved by varying the
energy of the state, holding all the parameters that define the
background fixed. In our discussion, the transition is achieved
by varying a parameter of the theory, the cosmological constant
$\Lambda$.

Another (related) difference between the two cases, is that in
\refs{\SusskindWS,\HorowitzNW} the matching between the entropies
of strings and black holes in the transition region can only be
done approximately, and both pictures suffer large corrections
there. In our case, the transition between the black hole and
string phases is sharp. For example, for $AdS_3$ one can define
an order parameter, $(c-c_{\rm eff})/c$, which vanishes in the
black hole phase, and is strictly positive in the perturbative
string one. Note that in the string phase, the difference between
$c_{\rm eff}$ and $c$ is not small, even in the semiclassical
limit $R_{AdS}\gg l_p$ \cbh. Indeed, for $k<1$ one has
\eqn\cceff{1-{c_{\rm eff}\over c}=\left(1-{1\over k}\right)^2~,}
so $c_{\rm eff}$ is a finite fraction of $c$ in the limit where
both go to infinity. Thus, one expects in general to be able to
see the difference between the two phases semiclassically.

The matching of the entropies of black holes and strings at the
transition point can be done precisely. We demonstrated that
the two agree for BTZ black holes with arbitrary mass and angular
momentum in $AdS_3$, and for black holes carrying up to two charges
in the linear dilaton case. The comparison was facilitated by the
fact that the black hole metric does not receive any $\alpha'$
corrections in these cases, so that the entropy formula obtained to
leading order in $\alpha'$ is in fact valid all the way to the
transition point, where the black hole abruptly dissappears due
to non-perturbative effects.

The resulting picture is also reminiscent of the phenomenon of
confinement in gauge theory. Fundamental strings behave like
partons; in the black hole phase, they are strongly interacting
at large distances, and form bound states -- the black holes. In
the string phase, their interactions go to zero at large distances
(near the boundary) and they behave like free partons in a
non-confining theory.

While the discussion of this paper was in the framework
of string theory, we expect the basic phenomena to be more
general. This leads to a natural question. Suppose we start
with three dimensional gravity with negative cosmological
constant, coupled to matter, with the action
\eqn\threedgrav{\CS={1\over 16\pi l_p}\int d^3x
\left(\CR+{2\over R_{AdS}^2}\right)+\CS_m~,}
where $\CS_m$ is the matter action. From the data in
\threedgrav\ one can construct the spacetime central
charge $c$ \cbh, but where is the information about
the phase the theory is in? In particular, if one is
not doing string theory, $l_s$ does not appear in
\threedgrav.

A likely answer to this question is the
following. Whether one is studying the action \threedgrav\
in the context of string theory or any other consistent
theory of quantum gravity, the theory will be in one of the
two phases we have described\foot{This should be true also
beyond the semiclassical approximation $R_{AdS}\gg l_p$.}.
The phase can be determined by computing the order parameter
$(c-c_{\rm eff})/c$ mentioned above. If the theory is in
a perturbative phase, with $c-c_{\rm eff}>0$,
it does not contain black holes as dynamical excitations
(\eg\ they cannot be formed by gravitational collapse of
normalizable states in the theory), and the high energy
density of states is dominated by the perturbative states
constructed from the fields in the Lagrangian \threedgrav.
If, on the other hand $c_{\rm eff}=c$, the theory is in
a non-perturbative phase and contains dynamical black holes.

An example of a theory of quantum gravity which does not
come from string theory and is in the perturbative phase
is Chern-Simons gravity \refs{\AchucarroVZ,\WittenHC}.
In this case, the spacetime CFT is Liouville theory
\CoussaertZP, or more precisely the conformal block of
the identity of any CFT with central charge $c$; it 
has\foot{In supersymmetric systems one finds instead
the superconformal block of the identity, for which 
$c_{\rm eff}$ is still of order one, and in particular, 
is much smaller than $c$ in the semiclassical regime.} 
$c_{\rm eff}=1$, and in 
particular $c_{\rm eff}<c$. We expect this theory not to 
contain dynamical BTZ black holes, just like two dimensional
string theory (or more generally, any asymptotically linear
dilaton theory with $Q>\sqrt2$) is not expected to contain 
$SL(2,\IR)\over U(1)$ black holes. There have been attempts 
in the past to calculate the entropy of BTZ black holes by 
studying three dimensional quantum gravity in the weakly 
coupled phase (for a recent review see \CarlipZN). From our 
general perspective it seems that these attempts are unlikely 
to succeed.

A particularly interesting special case of our discussion
is theories with $k=1$, which lie on the boundary between
the black hole and string phases. The fact that these theories
are special was noted already in \KutasovUA, but our discussion
clarified their significance -- in these theories one expects
black holes to be ``almost normalizable'' and to have the same
properties as highly excited fundamental strings. From the point
of view of the order parameter \cceff, the theories with $k=1$
look like critical theories at a second order phase transition.

In the linear dilaton case, an example of a theory with $k=1$
is the near-horizon geometry of the conifold
\eqn\connear{\IR^{3,1}\times\IR_\phi\times S^1~.}
If we compactify $\IR^{3,1}\to \IR^{1,1}\times T^2$ and
add $Q_1$ fundamental strings in the linear dilaton throat,
we get, at low energies, a CFT dual to string theory on
\eqn\adscon{AdS_3\times S^1\times T^2~,}
again with $k=1$. One sense in which this theory is critical
is that the slope of the linear dilaton on long strings
propagating near the boundary of $AdS_3$ \qqlloo\ vanishes
in this case. From the discussion of section 2 (see the
discussion around eq. \symprod), we expect the spacetime
CFT corresponding to \adscon\ to behave at high energies as
\eqn\stcft{\left(\IR\times S^1\times T^2\right)^{Q_1}/S_{Q_1}~.}
It is intriguing that this CFT closely resembles the
spacetime CFT corresponding to parallel fivebranes and strings
(whose near-horizon geometry is given by \simpads), if one
formally sets the number of fivebranes $k$ to one. This
does not really make sense in that context, since a single
fivebrane does not have a linear dilaton throat, but we see that
the CFT \stcft\ nevertheless makes an appearance, as the spacetime 
CFT corresponding to the critical, conifold, case.

The fact that the dilaton on long strings near the boundary
is constant in this case suggests that one can think of
\stcft\ as providing a light-cone description of a theory
that is Lorentz invariant in $5+1$ dimensions, at least at
high energies. It would be interesting to investigate
this possibility further.

Another interesting fact is that the backgrounds \connear,
\adscon\ corresponding to the critical case $k=1$ are $5+1$
dimensional. As we briefly reviewed earlier, asymptotically
linear dilaton and $AdS_3$ backgrounds of the sort studied
here should be thought of as theories of fivebranes wrapped
around various cycles, with or without fundamental strings.
It is possible that from that point of view, the transition
we have observed is due to the fact that for $k>1$, the
fivebrane is embedded in a space which is more than $5+1$
dimensional, and there is ``room'' for transverse fluctuations.
These fluctuations are described, at high energies, by black holes. 
For $k<1$, the fivebrane is embedded in a target space which is 
lower dimensional than its own worldvolume, and thus does not 
have transverse fluctuations. Hence it is in a different, 
perturbative, phase.

This is analogous to studying the behavior of strings in
a $d$ dimensional target space. As is well known, for $d>2$,
the strings have transverse fluctuations, which are reflected
in the appearance of a Hagedorn spectrum, whereas for $d<2$
there is no Hagedorn spectrum, and the theory is essentially
topological. The critical theory in that case is two
dimensional string theory.

It is natural to ask whether any of the phenomena that we have
found in $AdS_3$ can occur for $AdS_{d+1}$ spacetimes with $d>2$.
This is relevant for ideas \FidkowskiFC\ to study black hole
physics, whose natural home is at large $R_{AdS}/l_s$ (large
`t Hooft coupling in the dual super Yang-Mills theory for $d=4$)
by continuing them to small `t Hooft coupling.

For $AdS_3$, such a continuation is impossible for two
reasons. One is that the ratio $R_{AdS}/l_s$ is bounded from below
by a number of order one (\eg\ for the class of theories \adsthree,
$R_{AdS}/l_s\ge 2/3$). The second is that the black holes
cease being normalizable at $R_{AdS}/l_s=1$. Could similar things
happen for $d>2$?

At first sight the answer seems to be no. For $d=4$, it is expected
that $R_{AdS}/l_s$ can vary freely from zero to infinity, since the
`t Hooft coupling of the dual gauge theory has this property.
Also, in contrast to $AdS_3$, the conformally invariant vacuum of
quantum gravity on $AdS_5$ should exist as a normalizable state for
all values of $R_{AdS}$, or `t Hooft coupling. The fate of large
$AdS$ black holes for small $R_{AdS}$ is not clear, and it would be
interesting to investigate it further.

\vskip 1cm
\noindent{\bf Acknowledgments:}
We thank O. Aharony, T. Banks, M. Berkooz, S. Elitzur, D. Israel,
S. Sethi and S. Shenker for discussions. A.G. thanks the EFI at 
the University of Chicago, where most of this work was done, for
its warm hospitality. A.G. and E.R. thank the ISF and the EU grant 
MRTN-CT-2004-512194 for partial support. E.R. thanks the BSF for 
partial support. D.K. thanks the Weizmann Institute for hospitality, 
and the DOE for partial support. A.S. thanks the EFI and the Department 
of Physics at the University of Chicago for its warm hospitality, 
and the Kersten fellowship and Horowitz Foundation for partial support.

\appendix{A}{Determination of the sign of $Q_l$}

As mentioned in the text, in taking the square root
in \qlsq\ to determine the linear dilaton slope on
a long string one encounters a sign ambiguity. The
authors of \SeibergXZ\ showed that in spacetimes of
the form \simpads, which have $k>1$, the correct sign is
\qqlloo. Here we will show that for a class of models
with $k<1$ \qqlloo\ is reproduced as well. In fact,
the argument below is general, and can be used for all
$k$ to prove \qqlloo.

Consider string propagation in $AdS_3\times S^1\times {\cal M}_n$,
where ${\cal M}_n$ is an $N=2$ minimal model with
$c({\cal M}_n)=3-{6\over n}$. This model was discussed
in subsection 4.1 of \GiveonZM. Chiral operators in the
spacetime CFT are described by RR vertex operators
obtained by dressing chiral worldsheet operators
$V_{Q_V}$ in ${\cal M}_n$ with $R$-charge $Q_V$
by certain operators in $AdS_3\times S^1$ with
\eqn\jjj{j={k\over 2}(Q_V-1)~.}
Their scaling dimension in the spacetime CFT is
(see eq. (4.14) in \GiveonZM):
\eqn\chigkp{h_{\rm bottom}=j+\half~.}
As discussed in the text, the target space of the
CFT living on a long string is
\eqn\tsls{\IR_\phi\times S^1_Y\times {\cal M}_n~,}
with the $\phi$ dependent dilaton
\eqn\lindil{\Phi(\phi)=-{Q_l\over 2}\phi~.}
Our aim is to compute $Q_l$.

Chiral operators in the CFT on the long string are
obtained by dressing the chiral operators $V_{Q_V}$
in ${\cal M}_n$ with chiral operators in the $N=2$
Liouville CFT, $\IR_\phi\times S^1_Y$:
\eqn\chils{e^{\beta(\phi+iY)}V_{Q_V}~, \qquad \beta=Qj~, \qquad
Q=\sqrt{2\over k}~.}
The value of $\beta$ in \chils\ can be determined by
comparing the behavior of the operator \chils\ at large
$\phi$ to that of the operator corresponding to \jjj, \chigkp\
described in \GiveonZM. The latter goes at large $\phi$ like
$\exp(Qj\phi)$; hence the former must have the same behavior.
Comparing the scaling dimension of \chils\ to that of \chigkp\
one finds
\eqn\heqh{-\half\beta(\beta+Q_l)+{\beta^2\over 2}+{Q_V\over 2}
=h_{\rm bottom}~.}
Solving for $Q_l$ leads to
\eqn\qlql{Q_l=-(k-1)Q~.}

\appendix{B}{A check of the relation $h_w\geq \Delta_{\rm min}^{(w)}$}

In this appendix we show that, for $k<1$, perturbative string 
states in the sector with winding number $w$ have spacetime 
scaling dimensions which are bounded from below by 
$\Delta_{\rm min}^{(w)}=wQ_l^2/8$ \mindimww.

In string theory on $AdS_3$ there are two kinds of states:
those that belong to principal continuous series representations
of $SL(2,\IR)$, with
\eqn\jjbb{j=-\half+ip~,\qquad  p\in\IR ~,}
where $p$ is the momentum along the radial direction of
$AdS_3$, and those that belong to principal discrete series
representations, for which $j$ is real and obeys the unitarity
condition \unitcond.

Consider first the continuous states \jjbb. These only exist
in sectors with $w\not=0$, since for $w=0$ they correspond to
``bad'' tachyons and are projected out by the GSO projection.
In the sector with winding $w$, which we take to be positive,
as in section 2, their spacetime scaling dimensions \hhww\ can
be written as
\eqn\hwbb{h_w=\Delta_{\rm min}^{(w)}+{1\over 4}\left( w-{1\over w}\right)
\left( 2-{1\over k}\right)+{1\over w}\left({p^2\over k}+N\right)~,}
by using eqs. \jjbb\ and \mindimww. Hence, it is clear that
\eqn\hwgdbb{h_w\geq \Delta_{\rm min}^{(w)}~.}
Equality requires $w=1$, and $p=N=0$. Note that $N=0$ corresponds
to the fermionic string tachyon, and one might think that it should
be projected out by the chiral GSO projection. In fact, this is not
the case, since the GSO projection acts differently for even and
odd $w$ \ArgurioTB. For even $w$, the tachyon is projected out, while
for odd $w$ it is not. This is similar to the situation in the non-critical
superstring \KutasovUA, where tachyons with odd winding around the
$S^1$ in \throat\ survive the GSO projection.

Thus, we see that for $w=1$, the dimension $\Delta^{(w)}_{\rm min}$
lies at the bottom of a continuum of long string states. For $w>1$,
the single string states \hwbb\ satisfy $h_w> \Delta_{\rm min}^{(w)}$,
and the state with dimension $\Delta_{\rm min}^{(w)}$ is a multi-string
state consisting of $w$ long strings which do not interact near the
boundary of $AdS_3$, as explained in section 2.

Next we turn to principal discrete series states, focusing
for concretenes on backgrounds of the form
\eqn\minads{AdS_3\times S^1\times \CM_n~,}
where ${\cal M}_n$ is an $N=2$ minimal model with
$c({\cal M}_n)=3-{6\over n}$ (see section 2).

Consider first the sector with $w=1$. Principal discrete
series states with $w=1$ are equivalent by spectral flow
to short strings with $w=0$ \MaldacenaHW. Thus, we should
check that short string states satisfy
$h_{\rm short}\ge \Delta_{\rm min}^{(1)}$. The lowest
lying states in a sector with given R-charge are the chiral
ones, so it is enough to perform the check for those. The
dimensions of short string chiral operators in the spacetime
CFT corresponding to \minads\ were found in \GiveonZM\ to be
given by
\eqn\stchir{h_i={i+1\over 2(n+1)}, \;\;\;i=0,1,...,n-2~.}
They are indeed larger than
\eqn\deltaminone{\Delta_{\rm min}^{(1)}={(k-1)^2\over 4k}=
{1\over 4n(n+1)}~.}
Turning to sectors with $w>1$, we use the observation that
for any state with dimension $h_i$ in the sector with $w=1$,
there is one in the sector with winding number $w$, which
is related to it by spectral flow and has dimension \hwhone
\eqn\stscdi{h_i^{(w)}={h_i\over w}+{k\over 4}\left(w-{1\over w}\right)~.}
Since all $h_i$ satisfy $h_i>\Delta_{\rm min}^{(1)}$,
we have
\eqn\hbig{h_i^{(w)}>{\Delta_{\rm min}^{(1)}\over w}
+{k\over 4}\left(w-{1\over w}\right)=\Delta_{\rm min}^{(w)}+
{1\over 4}\left( w-{1\over w}\right)
\left( 2-{1\over k}\right)~.}
Hence, we conclude that these states satisfy \hwgdbb.

\appendix{C}{$SL(2,\IR)\times U(1)\over U(1)$}

In this appendix we briefly review, following
\refs{\GiveonGE,\ElitzurVW}, the gauging of the
$SL(2,\IR)\times U(1)$ WZW model by a family of anomaly
free U(1) subgroups and the manner by which it leads to the
rotating, charged black string \bssol. We emphasize those features of the
gauging procedure that are less standard. We then calculate
various properties of the black hole: its entropy, ADM mass
and two charges.

The background is constructed as follows \refs{\GiveonGE,\ElitzurVW}.
Let $(g,x)\in SL\(2,\IR\)\times U\(1\)$ be a point on the product
group manifold. The $U(1)$ gauge group is a subgroup of the 
$U(1)_L^2\times U(1)_R^2$ symmetry, which acts as
 \eqn\gatr{(g,x_L,x_R)\simeq (e^{\rho
\sigma_{3}}g e^{\tau \sigma_{3}},
 x_L+\rho',x_R+\tau')~,}
where $x_{L,R}$ are the left-moving and right-moving parts of $x$,
respectively. The right-moving part of the gauged $U(1)$ corresponds 
to $\vec\tau\equiv(\tau,\tau')=\tau\hat u$, where 
\eqn\uparam{\hat u=\(\matrix{\cos(\chi)\cr\sin(\chi)}\),}
is a fixed unit vector. The left-moving part corresponds to
$\vec\rho\equiv(\rho,\rho')=R\vec\tau$, where $R$ is the 
orthogonal $2\times 2$ matrix
 \eqn\orth{R=\(\matrix{\cos(\psi)&\sin(\psi)\cr
 -\sin(\psi)&\cos(\psi)}\)~.}
The fact that $|\vec\rho|=|\vec\tau|$ is necessary for 
an anomaly free gauging. 

The gauged WZW action takes the form
 \eqn\act{\eqalign{S=&{k\over {4\pi}}
[\int_{\Sigma} Tr(g^{-1}\partial g g^{-1}\bar \partial g)-{1
 \over 3} \int_{B} Tr (g^{-1}dg)^3 ]+{k\over {2 \pi}}\int_{\Sigma}\d x
\bar \d x\cr +&{k \over {2\pi}}\int
d^2z[ A\bar{\bf J}^T\hat u +\bar{A}{\bf J}^T R\hat u
 +2A\bar{A}\(R\hat u\)^T M \hat u]~,}}
where $k$ is the level of $SL(2,\IR)$,
$\Sigma$ is the worldsheet, $B$ is a three-manifold
whose boundary is $\Sigma$, and $(A,\bar A)$ is the $U(1)$ 
gauge field. The currents ${\bf J}^T$ and $\bar {\bf J}^T$ 
are the row vectors
 \eqn\cur{{\bf J}^T= (Tr[ \partial g
g^{-1}\sigma_3], 2\d x)~, \qquad \bar{\bf J}^T= (Tr[g^{-1}
\bar{\partial} g\sigma_3], 2\bar{ \partial} x)~.}
The $2\times 2$ matrix $M$ in \act\ is:
 \eqn\qfo{M=\left(\matrix{{1\over 2}Tr[g^{-1}\sigma_3 g\sigma_3]
 &0\cr 0&1}\right) + R~. }
The gauge field can be removed by a gaussian integration.
Following eqs. (14)--(23) in \GiveonGE,
after fixing the gauge we obtain the
worldsheet action and dilaton\foot{The coordinates
$\{\theta,y,x\}$ cover only one out of various regions of the
background; see \GiveonGE.}
 \eqn\lagrang{\eqalign{S=&{k\over
2\pi}\int d^2z\{\d\theta\d\bar\theta+\d x\bar\d x-\sh^2(\theta)\d
y\bar\d y\cr +&2\Delta^{-1}\[\sh^2(\theta)\cos(\chi)\bar\d
y+\sin(\chi)\bar\d x\]\[\sh^2(\theta)\cos(\chi-\psi)\d
 y-\sin(\chi-\psi)\d x\]\}~, \cr
 &e^{-2\Phi^{(3)}}=e^{-2\Phi_0^{(3)}}\Delta~,}}
where $\Phi_0^{(3)}$ is a constant and
 \eqn\DDelta{\Delta=1+\ch^2(\theta)\cos(\psi)+\sh^2(\theta)\cos(2\chi-\psi)~.}
It represents a three dimensional rotating black string with a non-trivial
antisymmetric $B$ field. To obtain the two charge two dimensional black hole, 
a KK reduction is done along the $x$ coordinate \GiveonGE; this results in a 
modified two dimensional metric and dilaton as well as two gauge fields, 
$A_G$ and $A_B$, whose origin is in the three dimensional metric and 
$B$ field,
 \eqn\kkk{\eqalign{g_{yy}=&-2k{\sh^2(\theta)\ch^2(\theta)
[1+\cos(\psi)][1+\cos(2\chi-\psi)]\over
 \Delta[1+\ch^2(\theta)\cos(2\chi-\psi)+\sh^2(\theta)\cos(\psi)]}~,\qquad
g_{\theta\theta}=2k~, \cr
 A_G=&{\sh^2(\theta)\sin(\psi)\over
 1+\ch^2(\theta)\cos(2\chi-\psi)+\sh^2(\theta)\cos(\psi)}dy~,\cr
 A_B=&2k{\sh^2(\theta)\sin(2\chi-\psi)\over \Delta}dy~,\cr
 e^{-2\Phi}=&e^{-2\Phi_0}
\sqrt{\Delta\(1+\ch^2(\theta)\cos(2\chi-\psi)+\sh^2(\theta)\cos(\psi)\)\over
 \(1+\cos(\psi)\)\(1+\cos(2\chi-\psi)\)}~,}}
where $\Phi_0$ is the value of the dilaton at the outer horizon $\theta=0$.
A string mode which is charged under $A_G$ ($A_B$) is a string mode
with $x$ momentum (winding).

Next we describe a few properties of the black hole. The ADM
mass is evaluated from \HawkingFD\
 \eqn\energy{M=-{1\over
 8\pi}\int_{\d\Sigma_t}\sqrt{-\sigma}
\[N^1K-N_0^{\,\,1}K_0-N^\mu P_{\mu\nu}r^\nu+
 e^{-\Phi}NB_{t\mu}H^{t\mu\nu}n_\nu\],}
where $\Sigma_t$ is a family of spacelike surfaces labeled by t,
$\d\Sigma_t$ is the boundary of $\Sigma_t$, $\sigma$ is the
determinant of the induced metric on $\d\Sigma_t$,
$P^{\mu\nu}$ is the canonical momenta conjugate to $G_{\mu\nu}$,
$N$ and $N^\mu$ are the lapse function and shift vector,
$n^\mu$ is the unit vector
normal to $\Sigma_\infty=\bigcup_t \d\Sigma_t$, $^1K$ is the one
dimensional extrinsic curvature on $\d\Sigma_t$, while $N_0$ and $^1K_0$
are the lapse function and the extrinsic curvature of the
reference background which, in our case,
is the pure linear dilaton background
 \eqn\pureld{\IR_\phi\times \IR_t\times S^1_x~,}
with $t\simeq y$ and $\phi\simeq\theta$.

One can use \energy\ in two ways. The first, by applying it
directly to the three dimensional background \lagrang\ in the
Einstein frame. The second, by adding a trivial fictitious $U(1)$
factor to the two dimensional string frame metric \kkk\ and then
passing to the Einstein frame in three dimensions\foot{The ADM
mass, as well as all other physical properties, are invariant
under a KK reduction which is trivial in the string frame.}.
This leads to the following expression for the mass:
 \eqn\ADMmass{M=-{1\over 2}\lim_{\theta\rightarrow\infty}
 \(\sqrt{|g_{tt}|\over g_{\theta\theta}}\d_\theta
 e^{-2\Phi}-\sqrt{2\over k}e^{-2\Phi}\)
=\sqrt{2\over k}{e^{-2\Phi_0}\over |\cos(\chi)+\cos(\psi-\chi)|}~,}
where $t\simeq y$ is normalized such that
$\lim_{\theta\rightarrow\infty}g_{tt}=1$.

The Bekenstein-Hawking entropy of \lagrang, \kkk\ is given
by\foot{The horizon area can be evaluated directly from the three
dimensional background \lagrang\ or by adding to the two dimensional 
one \kkk\ an extra circle in the string frame.}
 \eqn\Sbhbs{S_{\rm bh}=2\pi e^{-2\Phi_0}.}
The computation of the charges is an application of the Gauss law,
as in \McGuiganQP.
The charges are the integrals of $J^y$ over $\theta$, where
 \eqn\current{\eqalign{J_G^\mu
=&{1\over\sqrt{-2kg}}
\d_\nu\(\sqrt{-g}e^{-2\Phi}G_{xx}F_G^{\mu\nu}\)~,\cr
 J_B^\mu=&\sqrt{-{2k\over g}}\d_\nu\(
\sqrt{-g}e^{-2\Phi}G_{xx}^{-1}F_B^{\mu\nu}\)~.}}
The resulting charges are
 \eqn\charges{\eqalign{q_G
=&{1\over\sqrt{2k}}\lim_{\theta\rightarrow\infty}\ \sqrt{-g}
 \ e^{-2\Phi}G_{xx}F_G^{\theta y}=2M\sin({\psi\over 2})
\cos({\psi\over 2}-\chi)~,\cr
 q_B=&{\sqrt{2k}}\lim_{\theta\rightarrow\infty}\ \sqrt{-g}\
e^{-2\Phi}G_{xx}^{-1}F_B^{\theta y}
 =2M\sin({\psi\over 2}-\chi)\cos({\psi\over 2})~.}}
The right and left charges are given by
 \eqn\RLcharge{\eqalign{q_R=&\half\(q_G-q_B\)=M\sin(\chi)~,\cr
 q_L=&\half\(q_G+q_B\)=M\sin(\psi-\chi)~.}}
{}From eqs. \lagrang, \ADMmass, \RLcharge\ one can now obtain the three
dimensional background in Schwarzschild-like coordinates \bssol\ by the
change of variables
 \eqn\changvar{r=\sqrt{2\over k}e^{-2\Phi^{(3)}(\theta)}\ ,\qquad t\simeq y~,}
where $\Phi^{(3)}(\theta)$ is the function given in eqs. \lagrang, \DDelta.
Finally, from eqs. \ADMmass, \Sbhbs\ and \RLcharge,
one obtains the entropy \sbs.

\listrefs
\end